\documentclass{article}

\PassOptionsToPackage{numbers, compress}{natbib}
\usepackage[preprint]{neurips_2025}

\usepackage[utf8]{inputenc} 
\usepackage[T1]{fontenc}    
\usepackage{hyperref}       
\usepackage{url}            
\usepackage{booktabs}       
\usepackage{amsfonts}       
\usepackage{nicefrac}       
\usepackage{microtype}      
\usepackage{xcolor}         

\usepackage{graphicx}
\usepackage{listings}
\usepackage{amsmath}

\title{Galaxy Zoo Evo: 1 million human-annotated images of galaxies}

\usepackage{setspace}
\author{%
  Mike Walmsley\\
  University of Toronto\\
  \And
  Steven Bamford \\
  University of Nottingham \\
  \And
  Hugh Dickinson \\
  Open University\\
  \And
  Tobias G\'eron \\
  University of Toronto \\
  \And
  Alexander J. Gordon\\
  University of Edinburgh\\
  \And
  Annette M.~N.~Ferguson\\
  University of Edinburgh\\
  \And
  Lucy Fortson \\
  University of Minnesota \\
\And
Sandor Kruk \\
European Space Agency \\
\And
Natalie Lines\\
University of Portsmouth\\
\And
  Chris J.\ Lintott\\
  University of Oxford \\
      \And
  Karen L. Masters\\
  Haverford College \\
      \And
    Robert G.~ Mann\\
    University of Edinburgh\\
    \And
  James Pearson\\
  Open University \\
      \And
  Hayley Roberts\\
  University of Minnesota \\
      \And
      Anna M.M.\ Scaife\\
      University of Manchester\\
      \And
Stefan Schuldt\\
Università degli Studi di Milano\\
\And
Brooke Simmons\\
  University of Lancaster \\
        \And
  Rebecca Smethurst\\
  University of Oxford \\
        \And
Josh Speagle\\
University of Toronto\\
        \And
  Kyle Willett\\
  Amazon\\
}

\begin{document}

\maketitle

\begin{abstract}
We introduce Galaxy Zoo Evo, a labeled dataset for building and evaluating foundation models on images of galaxies. GZ Evo includes 104M crowdsourced labels for 823k images from four telescopes. Each image is labeled with a series of fine-grained questions and answers (e.g. `featured galaxy, two spiral arms, tightly wound, merging with another galaxy'). These detailed labels are useful for pretraining or finetuning. We also include four smaller sets of labels (167k galaxies in total) for downstream tasks of specific interest to astronomers, including finding strong lenses and describing galaxies from the new space telescope \textit{Euclid}. We hope GZ Evo will serve as a real-world benchmark for computer vision topics such as domain adaption (from terrestrial to astronomical, or between telescopes) or learning under uncertainty from crowdsourced labels. We also hope it will support a new generation of foundation models for astronomy; such models will be critical to future astronomers seeking to better understand our universe.
\end{abstract}

\section{Introduction}
\label{sec:introduction}

The development of computer vision methods has historically been driven by longstanding benchmarks like ImageNet \cite{deng_imagenet_2009}. But such benchmarks may no longer be good proxies for real-world performance \cite{fang_does_2023, tuggener_is_2022}.
Further, purpose-built benchmarks may (by design) elide important complexities of real-world tasks such as label uncertainty \cite{recht_imagenet_2019,Aitchison2021}. 
This paper aims to support new practical vision methods by sharing a large-scale crowdsourced dataset, created with scientific rigor, carefully documented, and free of privacy, safety, or commercial concerns, to serve as a challenging real-world benchmark.

\begin{figure}
    \centering
    \includegraphics[width=\linewidth]{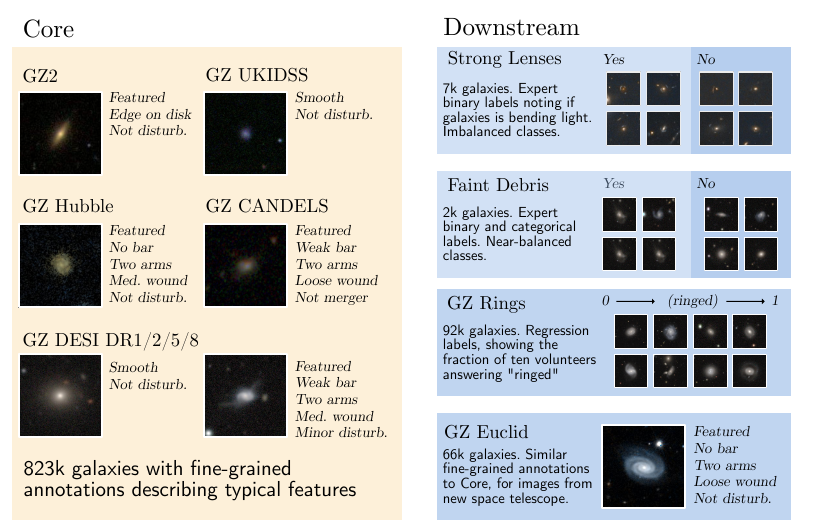}
    \caption{Overview of GZ Evo. GZ Evo includes a large-scale `Core' dataset with fine-grained annotations describing the typical features of 823k galaxies, and four smaller `Downstream' datasets with annotations of specific scientific interest.}
    \label{fig:dataset_overview}
\end{figure}

Galaxy Zoo Evo (Fig. \ref{fig:dataset_overview}) is made up of a `Core' dataset of galaxy images taken by four telescopes, where volunteers have answered a broad series of questions about the galaxy in each image, plus four smaller `Downstream' datasets of specific scientific interest. 

\begin{figure}
    \centering
    \includegraphics[width=\linewidth]{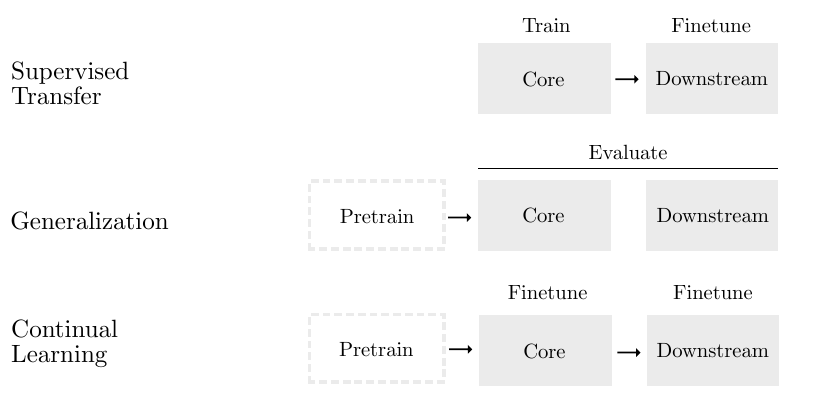}
    \caption{Three illustrative workflows using GZ Evo. Supervised transfer: train a supervised model on `Core' (galaxy images with general descriptions) and finetune on `Downstream'. Generalization evaluation: test an existing model on `Core' and `Downstream'. Continual learning: finetune an existing model on `Core', and then finetune again to maximise performance on `Downstream'}
    \label{fig:workflows}
\end{figure}

Figure \ref{fig:workflows} shows three illustrative workflows for using GZ Evo. 
One might train a supervised model on our Core dataset and then finetune it on our Downstream datasets (e.g. \cite{walmsley_scaling_2024}).
Or one might take an existing model and separately apply supervised finetuning to evaluate generalization (zero-shot similarity search, supervised finetuning, etc.) on each of our Core and Downstream datasets (e.g. \cite{Stein2022,smith_astropt_2024,parker_astroclip_2024,euclid_collaboration_euclid_2025}.
Or one might use continual learning  to finetune an existing model first on our Core datasets and then further finetune it on our Downstream datasets (see our SoViT-400m baseline, Sec. \ref{sec:downstream_results}).

Our Core dataset is composed of 823k labelled images of galaxies from the Galaxy Zoo project \cite{Masters2019a}. Galaxy Zoo asks members of the public to volunteer their time annotating images. These annotations -- and models trained on these annotations -- help us learn how galaxies evolve; measuring the appearance of millions of galaxies, each with a different history, allows one to identify how galaxies grow, merge, and age. \cite{Buta2013}.
We collate sixteen years of annotations by 334k volunteers.

Our Downstream datasets are composed of four smaller sets of images with labels of specific scientific interest. Three of the Downstream datasets are searches for a particular type of galaxy:  strong lenses (where the light from a background galaxy is bent by the gravity of a foreground galaxy); galaxies with rings; and galaxies with faint debris (often indicating a recent collision with another galaxy). The fourth dataset is GZ Euclid, where volunteers answered the same series of broad questions as Core but for images from a new and state-of-the-art space telescope, \textit{Euclid}. Generalizing to images from \textit{Euclid} is a real and urgent test for foundation models in astronomy.

No astronomy knowledge is required. We include minimal and reproducible code for training and evaluating baseline models.
Our framework is extensible to new data, allowing Galaxy Zoo `Evo' to `evolve' as new telescopes and new labels become available. 
\\
\\
Galaxy Zoo Evo is available at \href{https://huggingface.co/mwalmsley}{huggingface.co/mwalmsley} with supporting code at \href{https://github.com/mwalmsley/gz-evo}{github.com/mwalmsley/gz-evo}.


\section{Related Work}
\label{sec:related_work}

This paper is inspired by the 2014 `Galaxy Challenge' Kaggle competition which asked competitors to predict volunteer votes to an early Galaxy Zoo iteration (GZ2, below). The winning submission \cite{Dieleman2015} introduced CNNs to astronomy and sparked a transition to deep learning methods throughout the discipline (see \cite{HuertasCompany2022} for a review). The underlying dataset of 62k training images has also been widely used in computer science as a real-world benchmark (see e.g. \cite{Katebi2019a,andrew_galaxy_2023,li_galaxy_2023,pandya_e2_2023}). Galaxy Zoo Evo is intended as a ten year update on the Galaxy Challenge dataset, bringing an order-of-magnitude increase in scale, a new focus on multi-task learning and and downstream generalization, and an extensible framework to allow the dataset to grow over the next ten years.

We are also motivated by Fang et al's call that `researchers should explicitly search for methods that improve accuracy on real-world non-web-scraped datasets, rather than assuming that methods that improve accuracy on ImageNet will provide meaningful improvements on real-world datasets as well' \cite{fang_does_2023}. This is especially relevant for domains with a substantial shift in features vs. ImageNet such as medical imaging\cite{raghu_transfusion_2019,bressem_comparing_2020,ke_chextransfer_2021} and manufacturing \cite{kasempakdeepong_sugarcane_2022,mezher_novel_2023}. Recent progress in vision-language pretraining, e.g., training models on extreme-scale web-scraped images, shows impressive few-shot gains on other web-scraped tasks \cite{Radford2019, zhai_scaling_2022, Agrawal2022VisionLanguagePC}. We hope Galaxy Zoo Evo will help support a growing theme of work investigating vision-language pretraining for improved performance on scientific and technical images (e.g. \cite{Agrawal2022VisionLanguagePC, Xiao2024ASIMSAAS,Yan2025MAKEMK,Liu2023ImprovingMV,Cho2024PretrainingVM,AtFirstSight2024}).

We build on a series of papers by the Galaxy Zoo collaboration (\cite{Walmsley2022decals,Walmsley2023desi,Walmsley2023zoobot,walmsley_scaling_2024}) that cumulatively motivate the design of our baselines. Our labelled dataset complements recent work to create an unlabelled multi-modal (i.e. images, time series, spectra) dataset for astronomy \cite{angeloudi_multimodal_2024}; together, these datasets form a comprehensive resource for self-supervised pretraining, supervised finetuning, and downstream evaluation. 

\section{Core Dataset}
\label{sec:galaxy_zoo_overview}

\subsection{Context and Construction}

Galaxy Zoo asks members of the public to label galaxy images. Volunteers are presented with a galaxy image and initial question. 
As they answer, they are led down a decision tree where the next question asked depends on the previous answer. This breaks down the process of identifying detailed galaxy features into a series of intuitive judgements. 
Volunteers are guided by a tutorial, illustrative icons, expandable help text, and a comprehensive `Field Guide'.
They can share discoveries and compare notes on an integrated forum.
We invite the reader to take part at \href{www.galaxyzoo.org}{www.galaxyzoo.org}.

The questions and images both change over time, reflecting advances in telescopes and scientific knowledge.
Galaxy Zoo publications therefore divide sets of questions and images into homogeneous labelling `campaigns'. 
Adjusting a question or sourcing images from a new telescope is considered the start of a new campaign.
The labels from each campaign were published separately, in disparate formats, in astronomical journals.
Galaxy Zoo Evo brings together the separate Galaxy Zoo campaigns in a consistent and machine-learning-friendly format.
We include all\footnote{Excluding the first campaign, Galaxy Zoo 1, which predates the decision tree and asked only one question} published Galaxy Zoo campaigns, covering four telescopes and a wide diversity of galaxies.
Table \ref{tab:subsets} shares key statistics for each campaign. The decision trees asked in each campaign are visualised at \href{https://data.galaxyzoo.org/gz_trees/gz_trees.html}{https://data.galaxyzoo.org/gz\_trees/gz\_trees.html}. 

\begin{table}
  \caption{Subset sizes. GZ Evo size is slightly smaller than naive sum to avoid train/test contamination from galaxies that appear in multiple campaigns. Logged-in volunteers are unique per subset.}
  \label{tab:subsets}
  \centering
  \begin{tabular}{llllll}
    \toprule
    Name     & Galaxies     & Logged-in Volunteers & Votes & Start Year & Data Release \\
    \midrule
    GZ2 & 209k  & 84k & 35.8M  & 2009 & \cite{Willett2013}   \\
    GZ Hubble     & 96k & 88k & 17.6M  & 2010 & \cite{Willett2017a}   \\
    GZ UKIDSS     & 71k  &   25k   & 9.5M & 2013 & \cite{masters_galaxy_2024} \\
    GZ CANDELS     & 48k   &  45k &  5.2M & 2012 & \cite{Simmons2017} \\
    GZ DESI     & 399k   &  125k &  39.1M & 2015 & \cite{Walmsley2023desi}\\
    \midrule
    GZ Evo     & 823k   &  334k &  104M & - & (here) \\
    \bottomrule
  \end{tabular}
\end{table}

\subsection{Labels, Loss Functions, and Metrics}

Each galaxy image in the Core dataset is annotated in a format like:

\begin{verbatim}
    {
        smooth-or-featured_gz2_smooth: 30
        smooth-or-featured_gz2_smooth_fraction: 0.75
        smooth-or-featured_gz2_featured-or-disk: 10
        smooth-or-featured_gz2_featured-or-disk_fraction: 0.25
        ...
        summary: `smooth'
    }
\end{verbatim}

The number of volunteers voting for an answer to a given question is reported in the columns \texttt{\{question\}\_\{campaign\}\_\{answer\}}. The total number of volunteers answering each question varies (e.g., because fewer volunteers reach questions further down the decision tree). Training directly on these counts using a multinomial loss\footnote{i.e., for each question, predict the odds $\hat{P}=f_\theta(x)$ that each volunteer selects each answer, and penalize the model according to the likelihood that we therefore observe $y=\Vec{k}$ volunteers selecting each answer} is a simple approach for incorporating label uncertainty. We train our baselines on Core using a multinomial loss summed across all questions and all campaigns (multi-task learning). We also report RMSE metrics\footnote{We precalculate the fraction of volunteers giving each answer to each question in the columns \texttt{\{question\}\_\{campaign\}\_\{answer\}\_\{fraction\}}}.

\begin{figure}
    \includegraphics[width=0.5\linewidth]{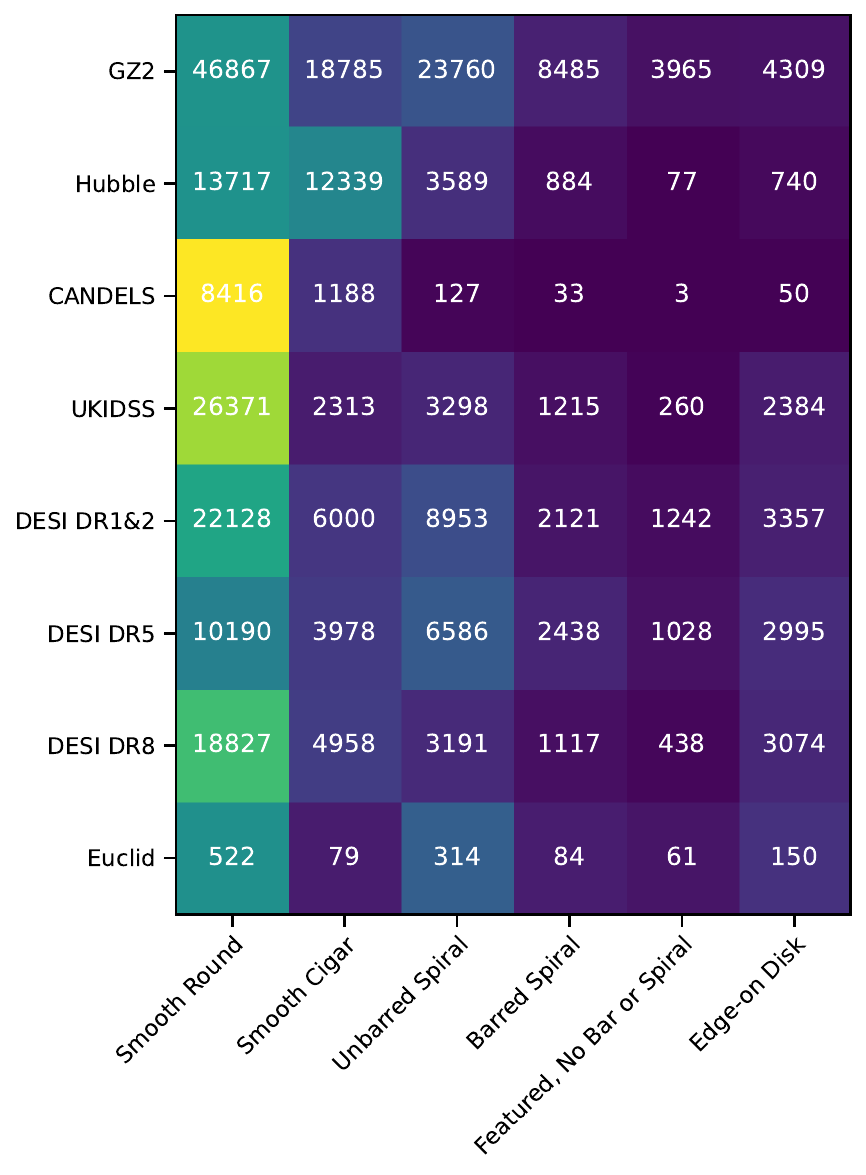}
    \hfill
    \includegraphics[width=0.5\linewidth]{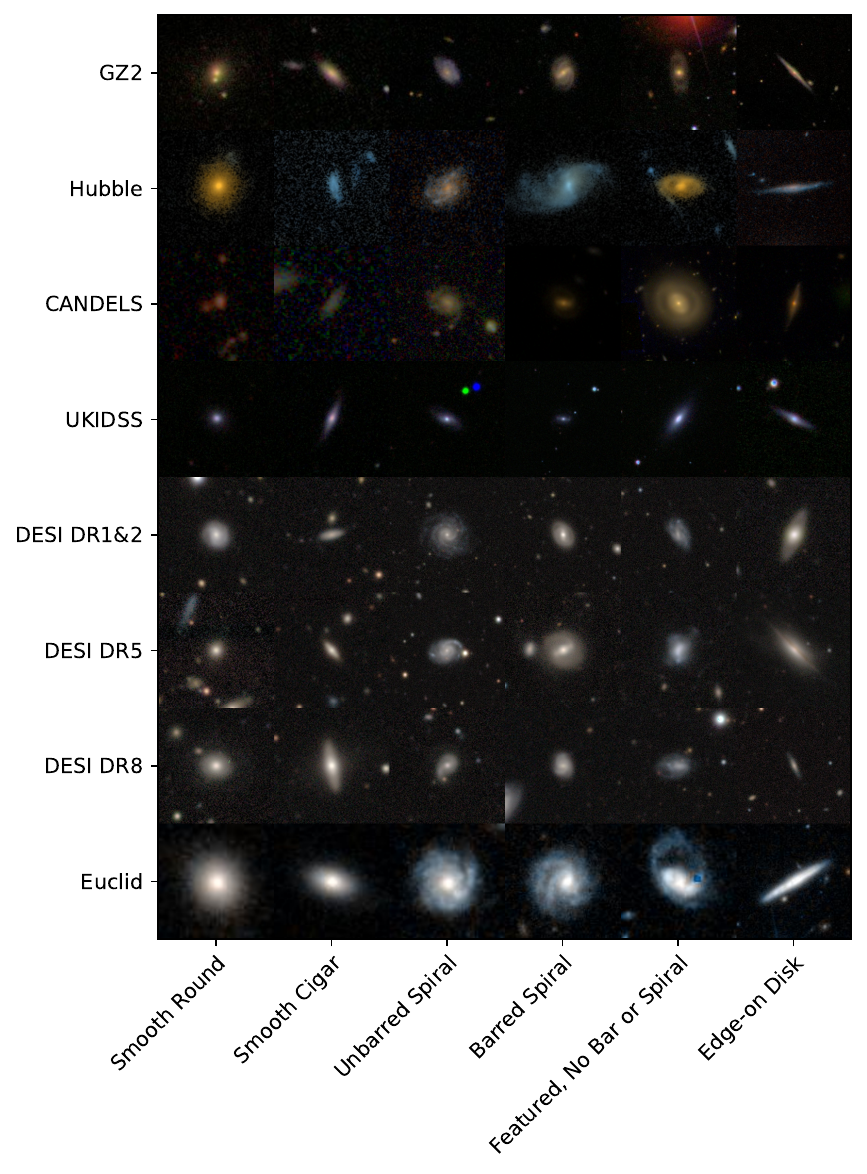}
        \caption{Left: Class counts by subset, when using our aggregated high-confidence class labels. Colours normalized by subset (row) to show the varying class distribution within each subset. This is a real(istic) test scenario for domain adaption. Right: an example galaxy image for each survey and class.}

        \label{fig:class_label_distributions}
\end{figure}

For researchers who prefer to work in a classification context, we also aggregate our count labels into six non-overlapping classes.
The class label is reported in the \texttt{summary} column.
Galaxies are labelled as smooth and round, smooth and cigar-shaped, edge-on disks, featured with spiral arms and a bar, featured with spiral arms and no bar, or featured with another characteristic (i.e. neither spiral arms nor a bar). 
We only provide a class label for the subset (287k) of galaxies where a class label can be confidently assigned. We consider as high confidence galaxies where a single answer received over half of all volunteer votes (an absolute majority), for all of the questions needed to derive our summary label, and where the galaxy received at least 30 volunteer votes in total (20 for GZ Euclid). We include code for reproducing our aggregation of votes to class labels. Figure \ref{fig:class_label_distributions} shows the distribution of classes in each subset and an example of every class for every campaign.

We anticipate the class labels will be useful for methods that assume a classification context (common in e.g. active learning, domain adaption) and where label uncertainty is not the research focus. We anticipate the count labels will be useful as a challenging multi-task finegrained learning problem, for pretraining and finetuning (Fig. \ref{fig:workflows}), and for learning under realistic uncertainty from crowds.

\section{Downstream Datasets}

We complement our Core dataset -- galaxy images with finegrained labels from volunteers -- with four downstream datasets intended for evaluating the generalization of trained foundation models.
Three downstream datasets (Strong Lenses, GZ Rings, and Faint Debris) ask models to answer a new task using images already seen. The remaining dataset (GZ Euclid) asks models to answer a comparable set of questions using images from the new \textit{Euclid} space telescope. 

\subsection{GZ Euclid}

The \textit{Euclid} space telescope is beginning to take images of one third of the sky. When compared to images from the ground, \textit{Euclid}'s images are approximately 10 times sharper\footnote{Astronomers measure image sharpness using the `point spread function full-width-half-maximum', which is calculated as the angular distance by which the apparent light recorded from a point-like object falls by half} and promise a revolution in our understanding of distant galaxies \cite{Q1-TP001}.
Galaxy Zoo volunteers were asked to annotate these new images using the same tree structure as for previous campaigns extensively annotated in Core \cite{Q1-SP047}. This downstream task requires the model to predict those volunteer responses. 
This directly tests the ability of models to generalise from GZ Evo Core to new telescopes. Additionally, because \textit{Euclid} is new, training galaxies only have 5 volunteer votes (test galaxies have 25) and so models must generalise from noisy training labels.

\subsection{Strong Lenses}

Mass bends light. When two galaxies line up, the foreground galaxy can bend and magnify the light of the background galaxy into striking visual arcs. This rare alignment allows astronomers to measure the mass of the invisible dark matter in the foreground galaxy, and, potentially, to measure the rate of the expansion of the universe \cite{shajib_lens_review}. This downstream task uses expert labels from a recent search for strong lenses in new \textit{Euclid} images \cite{Q1-SP048, Q1-SP052, Q1-SP053, Q1-SP054, Q1-SP059} and requires models to predict whether 10 experts judged each image to contain a strong lens or a similar-looking `imposter' galaxy. Strong lenses are rare and hence the dataset is highly imbalanced (6.7\% lens).

\subsection{GZ Rings}

Ringed galaxies have (as one expects) a circular feature within or around the galaxy. Their formation mechanism is controversial \cite{Buta2017catalog}. To identify ringed galaxies, we use labels collected via Galaxy Zoo Mobile over two years. Volunteers were shown a galaxy image and asked to swipe according to whether it included a ring. This downstream task requires models to predict the fraction of ten volunteers who responded `Ring'.

\subsection{Faint Debris}

Galaxies grow partly by colliding and merging with one another. These collisions leave faint but visible remnants: `low surface brightness features', in astronomy jargon, or `faint debris' here. Identifying these remnants allows astronomers to measure the rate of galaxy growth, but remnants are hard to distinguish automatically \cite{Pawlik2016, Bottrell2019, Pearson2019Mergers}.
\citealt{gordon_uncovering_2024} annotated 2000 images from DECaLS \cite{Dey2018, Walmsley2022decals} according to whether each image showed faint debris and, if so, assigned a subclass. Subclasses are `arm', `stream', `shell', or `diffuse' \cite{Atkinson2013}. We use these labels for two downstream tasks: predict if the image has debris (`Is Debris') and (where positive) the subclass (`Which Debris').

\section{Core Dataset Results}
\label{sec:core_results}
\label{app:recipe}

We train popular architectures (ConvNeXt \cite{Liu2022}, EfficientNetv2 \cite{Tan2021}, MaxViT \cite{Tu2022}, ResNet \cite{He2016a, wightman_resnet_2021}, SoVit-400m \cite{alabdulmohsin_getting_2024}) with a simple standard recipe: schedule-free AdamW with weight decay (0.05), \cite{loshchilov_decoupled_2019}, stochastic depth \cite{huang_deep_2016} (values per the original papers), and standard rotations/flips/crops for augmentations. We select architecture variants from 10 A100-hours (ConvNext-Nano) to 80 A100-hours (SoViT-400m/14). We use a learning rate of $10^{-4}$, reducing to $10^{-5}$ for the largest models. 

Table \ref{tab:core_metrics} reports performance metrics on our Core Galaxy Zoo subsets. We report classification metrics for models trained with cross-entropy loss against our class labels, and regression metrics for models trained with multinomial loss against our vote counts (`k of N volunteers'). 
No single architecture is clearly strongest; EfficientNetV2 and Max-ViT are strong at classification and multi-task regression, respectively, while ConvNeXt is strong overall. The largest models perform poorly due to overfitting. All modern architectures significantly outperform ResNet50.

Regression metrics show a relatively small spread vs.~other datasets because many answers have similar vote fractions for most galaxies. For example, due to physical symmetries, galaxies typically have 0, 2, or 4 spiral arms, and very few have 1 or 3. Any model will achieve a low RMSE/loss at predicting the fraction of volunteers voting for 1 or 3 spiral arms (almost none), and this reduces the variation in RMSE/loss when averaging across all answers. This does not imply that predicting vote fractions for all answers is easy to solve. 
Below, we finetune the regression-trained models on our Downstream datasets, and show they outperform the equivalent ImageNet-trained models.

\begin{table}
  \caption{Metrics for supervised baselines training on GZ Evo Core dataset. }
  \label{tab:core_metrics}
  \centering
  \begin{tabular}{lrrrrr}
    \toprule
    &
    \multicolumn{3}{c}{Classification} & \multicolumn{2}{c}{Regression} \\
    \cmidrule(lr){2-4}
    \cmidrule(lr){5-6}
    \toprule
    Architecture     & Accuracy     & Balanced Acc. & CE Loss & RMSE & Multinomial Loss \\
    \midrule
    ResNet50 & 0.8872	 & 0.719 & 1.1961 & 0.0960 & 17.602   \\ %
    ConvNeXT Nano & 0.8791 & 0.595 & 1.7019 & 0.0874 & 17.173 \\ 
    ConvNeXT Base & 0.9300 & \textbf{0.813} & 0.6863 & \textbf{0.0842} & 16.967 \\
    ConvNeXT Large & 0.8936 & 0.612 & 1.3430 & 0.0864 & 17.271 \\
    EfficientNetV2 S & \textbf{0.9310} & 0.810 & \textbf{0.6222} & 0.0878 & 17.150 \\ 
    EfficientNetV2 M & 0.9220 & 0.784 & 0.6837 & 0.0879 & 17.100 \\
    EfficientNetV2 L & 0.9259 & 0.799 & 0.6237 & 0.0869 & 17.105 \\
    MaX-ViT Tiny & 0.9275 & 0.804 & 0.7639 & 0.0858 & \textbf{16.965} \\
    MaX-ViT Base & 0.9013 & 0.696 & 0.9237 & 0.0852 & 17.004 \\
    \bottomrule
  \end{tabular}
\end{table}

\subsection{Downstream Dataset Results}
\label{sec:downstream_results}

Table \ref{tab:downstream_test_loss} shows test loss when training on the GZ Evo Core dataset and then finetuning on each of the GZ Evo Downstream datasets (mean of six seeds). For finetuning, we adapt all layers with a weight decay of 0.2 and an aggressive\footnote{With the exception of SoViT-400m, where we use $L_{i-1} = L_{i}^{0.8}$ due to the atypically large number of blocks} exponential learning rate reduction by block (layer group) of $L_{i-1} = L_{i}^{0.2}$. In addition to test loss by dataset, we report an \textit{overall score} calculated by rescaling the test loss for each dataset from 0 (worst model) to 1 (best) and then taking a mean across datasets. 

Overall, ConvNeXt models perform strongest downstream, closely followed by MaxViT models. SoViT-400m/14 excels at identifying strong lenses, possibly because identifying multiple separated arcs of lensed light benefits from an architecture designed for nonlocal features. ConvNeXt-Nano performs best at the smallest two datasets (Is Debris and Which Debris) while ConvNeXt-Base performs best at the two largest Downstream datasets (GZ Euclid, GZ Rings). 

For comparison, we also include several models directly-finetuned from their non-domain-specific pretrained versions (ImageNet-1k \cite{deng_imagenet_2009} for ResNet50, ImageNet-12k\footnote{\url{https://github.com/rwightman/imagenet-12k}} for ConvNeXt-Nano, and webli \cite{chenPaLIJointlyScaledMultilingual2023} via SigLIP 2 \cite{tschannenSigLIP2Multilingual2025} for SoViT-400m). These perform worse in every case, suggesting a continual learning approach is useful for maximising downstream performance. Notably, when considering only the three models finetuned directly on our downstream datasets without intermediate Core training, the SigLIP-trained model (SoViT-400m) is more succesful than the ImageNet-1k/12k models (ResNet/ConvNexT), which may suggest that features from SigLIP (i.e. vision-language training) may generalise better to our astronomy images than features learned from supervised ImageNet, consistent with similar observations in other domains \cite{Yan2025MAKEMK}.

\begin{figure}
    \centering
    \includegraphics[width=\linewidth]{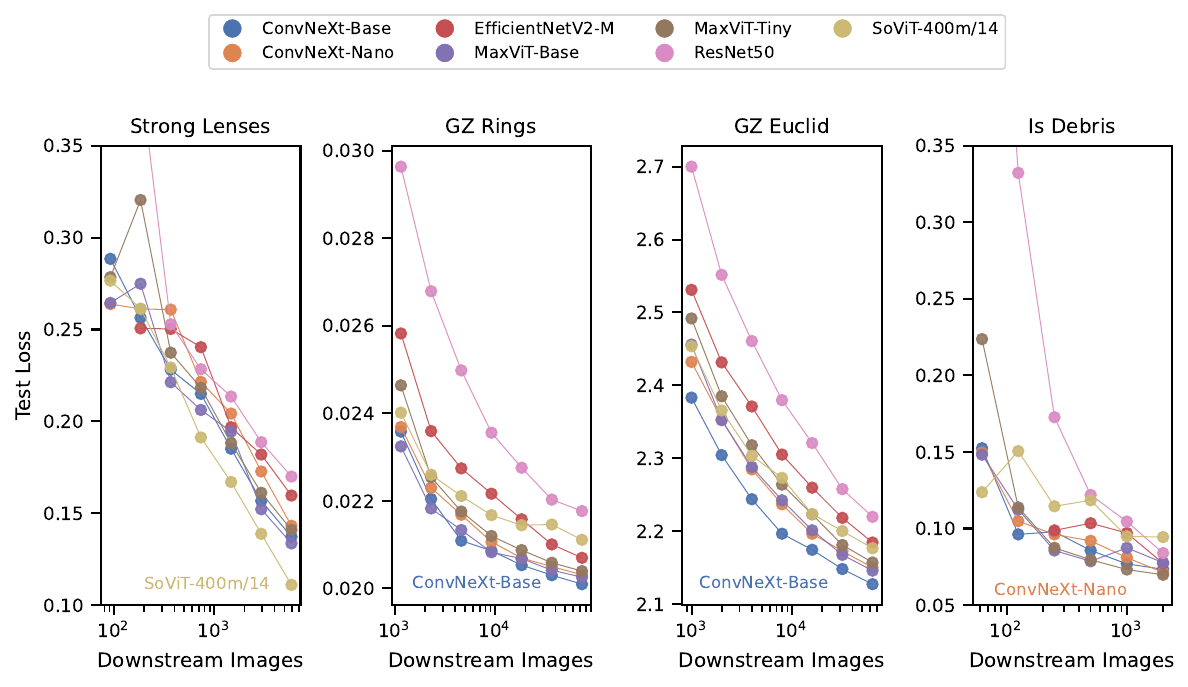}
    \caption{Test loss vs.~dataset size when training on the GZ Evo Core dataset and then finetuning on the GZ Evo Downstream datasets. Mean of six seeds. ConvNeXt-Base performs best at GZ Rings, GZ Euclid, and overall. SoViT-400m/14 performs best at finding strong lenses. Small modern models (MaxViT-Tiny and ConvNeXt-Nano) perform best at our smaller Is Debris dataset.}
    \label{fig:finetuning_vs_dataset_size}
\end{figure}

Figure \ref{fig:finetuning_vs_dataset_size} investigates label-efficient learning. We again show test loss when training on Core and then finetuning on each Downstream dataset, but artificially restrict the size of each Downstream dataset to include fewer images for finetuning. Consistent with Table \ref{tab:downstream_test_loss}, all models tested outperform their directly-finetuned equivalents at all dataset sizes (not shown for clarity). All models show consistent and predictable improvement when scaling dataset size, suggesting that, first, data is a bottleneck to scaling, and second, more efficient strategies to learn from limited data (or from unlabelled data) should improve on our baselines.

\begin{table}
    \caption{Test loss when training on the GZ Evo Core dataset and then finetuning on the GZ Evo Downstream datasets. Overall score is normalised from 1 (best model at every dataset) to 0 (worst model at every dataset). Mean of six seeds. `*' indicates otherwise-identical models without intermediate training on GZ Evo Core; these perform worse in every case. SoViT-400m/14 performs best for strong lenses, while ConvNeXt models are strongest overall.}
    \label{tab:downstream_test_loss}
    \centering
\begin{tabular}{lrrrrrr}
\toprule
 & Strong Lenses & GZ Euclid & GZ Rings & Is Debris & Which Debris & Overall Score \\
Model &  &  &  &  &  &  \\
\midrule

ConvNeXt-Base & 0.1372 & \textbf{2.1271} & \textbf{0.0201} & 0.0734 & 0.6709 & \textbf{0.97} \\
ConvNeXt-Nano & 0.1431 & 2.1500 & 0.0203 & 0.0713 & \textbf{0.6334} & 0.95 \\
MaxViT-Base & 0.1336 & 2.1457 & 0.0203 & 0.0777 & 0.6918 & 0.93 \\
MaxViT-Tiny & 0.1409 & 2.1568 & 0.0204 & 0.0698 & 0.6631 & 0.93 \\
ViT-Base/16 & 0.1609 & 2.1476 & 0.0205 & 0.0673 & 0.6614 & 0.92 \\
Beit3-Base & 0.1783 & 2.1653 & 0.0208 & 0.0755 & 0.6438 & 0.88 \\
ConvNeXtV2-Base & 0.1687 & 2.1572 & 0.0211 & \textbf{0.0656} & 0.7469 & 0.86 \\
SoViT-400m/14 & \textbf{0.1232} & 2.1883 & 0.0211 & 0.1002 & 0.7050 & 0.85 \\
EfficientNetV2-M & 0.1596 & 2.1843 & 0.0207 & 0.0773 & 0.7511 & 0.84 \\
SoViT-400m/14* & 0.1309 & 2.1921 & 0.0216 & 0.0974 & 0.8088 & 0.79 \\
EfficientFormerV2-L & 0.3507 & 2.1451 & 0.0209 & 0.0904 & 0.6766 & 0.72 \\
ResNet50 & 0.1699 & 2.2195 & 0.0218 & 0.0839 & 0.9118 & 0.70 \\
ConvNeXt-Nano* & 0.1720 & 2.2251 & 0.0223 & 0.1291 & 0.9879 & 0.61 \\
ResNet50* & 0.2463 & 2.3208 & 0.0274 & 0.2811 & 1.1844 & 0.09 \\

\bottomrule
\end{tabular}

\end{table}

\section{Limitations}
\label{sec:limitations}

GZ Core, GZ Rings, and GZ Euclid are annotated by volunteers, and Strong Lensing is annotated by professional astronomers. Volunteer annotations of galaxy images have been consistently shown to closely approximate professional annotations. However, for both volunteer and professional annotators, inter-annotator disagreement is expected.  See Appendix \ref{app:volunteer_and_professional}. Real scientific data often cannot be annotated perfectly by humans and operating under ambiguity is necessary for real-world machine learning systems.

For annotations using a decision tree structure (Core and GZ Euclid), the number of volunteers answering each question varies. Fine-grained questions deeper in the tree collect fewer votes\footnote{For example, if we ask 40 people `Is this galaxy smooth or featured?', and 25 answer `Featured', then 15 further select `Spiral', only those 15 people would be asked `How many spiral arms?'}. This introduces heteroskedastic uncertainty and motivates our choice to share only high-confidence labels for our classification setting. The benefit of a decision tree is that annotators are only asked relevant questions; we hope our decision tree labels will support research into efficient crowdsourcing where finegrained labels are necessary but annotation is expensive (e.g. medical imaging).

We broadly find that the largest variants of the models tested tend to underperform their moderate-size variants, particularly for our smaller datasets. In principle, large variants should at least match the performance of small variants (e.g. by trivially replicating the small variant) and, where data is plentiful, are often shown to exceed the performance of smaller models. In practice, properly regularizing large models on small datasets is nontrivial, and our finetuning recipe likely causes these large models to overfit. Evaluating the effectiveness of new approaches for few-shot finetuning of large models is a key goal of GZ Evo.

GZ Evo is not suitable for research requiring raw telescope data. GZ Evo provides RGB images to support researchers without an astronomy background.

\section{Data Access}
\label{sec:data_access}

Data is available via the HuggingFace Hub. You can download them with e.g. 
\begin{lstlisting}[language=Python, basicstyle=\ttfamily\footnotesize, keywordstyle=\color{blue}, commentstyle=\color{green!50!black}, stringstyle=\color{red}]
from datasets import load_dataset
dataset = load_dataset("mwalmsley/gz_evo", split="train")
# your framework of choice e.g. numpy, tensorflow, jax, etc
dataset.set_format("torch")  
\end{lstlisting}

\section{Conclusion}
\label{sec:conclusion}

Galaxy Zoo Evo brings together sixteen years of effort to annotate one million images of galaxies. 
GZ Evo is large-scale, meticulously collected, and free of privacy, safety, or commercial concerns. It is designed for building and evaluating foundation models, and also suits broader research into practical methods for
crowdsourcing, label uncertainty, multi-task learning, and active learning.

GZ Evo is a living dataset. New telescopes drive new discoveries;
our modular multi-task format makes it straightforward to add new labelled galaxy images from any new telescopes.
Researchers can then seamlessly apply the same training code to new GZ Evo versions, creating new telescope-ready finetuned models in weeks instead of years.

New telescope data is also ideal for testing the generalisation of foundation models. 
It is difficult to cleanly evaluate such models because any internet data, including standard benchmarks, is in-distribution (or worse, memorized training data).
We know for certain that current foundation models could not have been trained on data from new telescopes, regardless of engineering resources, because the data did not previously exist. 
Generalization performance to new telescopes, as measured with GZ Evo, is an ironclad test for existing models.

\newpage

\begin{ack}

The data in this paper are the result of the efforts of the Galaxy Zoo volunteers, without whom none of this work would be possible. Their efforts are individually acknowledged at \href{http://authors.galaxyzoo.org}{http://authors.galaxyzoo.org}. We would also like to thank our volunteer translators; Mei-Yin Chou, Antonia Fernández Figueroa, Rodrigo Freitas, Na'ama Hallakoun, Lauren Huang, Alvaro Menduina, Beatriz Mingo, Verónica Motta, João Retrê, and Erik Rosenberg.

This publication uses data generated via the Zooniverse.org platform, development of which is funded by generous support, including a Global Impact Award from Google, and by a grant from the Alfred P. Sloan Foundation.

This research made use of the open-source Python scientific computing ecosystem, including SciPy \citep{virtanen_scipy_2020}, Matplotlib \citep{Hunter2007}, scikit-learn \citep{Pedregosa2012}, scikit-image \citep{VanderWalt2014} and Pandas \citep{McKinney2010}. This research made use of Astropy, a community-developed core Python package for Astronomy \citep{TheAstropyCollaboration2018}. This research made use of PyTorch \citep{Paszke2019} and Pyro \citep{bingham2018pyro}.

MW is a Dunlap Fellow. The Dunlap Institute is funded through an endowment established by the David Dunlap family and the University of Toronto. LF, HR and KBM acknowledge partial support from the US National Science Foundation under grants OAC 1835530 and IIS 2006894. BDS acknowledges support through a UK Research and Innovation Future Leaders Fellowship [grant number MR/T044136/1].  CJL acknowledges funding from the Alfred P. Sloan foundation.

This research was enabled in part by support provided by the University of Toronto and the Digital Research Alliance of Canada (alliance.can.ca).

We would like to thank Jo Bovy, Micah Bowles, Nolan Koblischke, and Devina Mohan for helpful discussions.

\end{ack}

{
\small
\bibliography{main}

@String(ICLR  = {Int. Conf. Learn. Represent.})

@String(ICLR  = {ICLR})

@article{Radford2019,
	title = {Learning {Transferable} {Visual} {Models} {From} {Natural} {Language} {Supervision}},
	url = {http://arxiv.org/abs/2103.00020},
	journal = {OpenAI},
	author = {Radford, Alec and Kim, Jong Wook and Hallacy, Chris and Ramesh, Aditya and Goh, Gabriel and Agarwal, Sandhini and Sastry, Girish and Askell, Amanda and Mishkin, Pamela and Clark, Jack and Krueger, Gretchen and Sutskever, Ilya},
	year = {2021},
	pages = {47},
}

@misc{alabdulmohsin_getting_2024,
	title = {Getting {ViT} in {Shape}: {Scaling} {Laws} for {Compute}-{Optimal} {Model} {Design}},
	shorttitle = {Getting {ViT} in {Shape}},
	url = {http://arxiv.org/abs/2305.13035},
	doi = {10.48550/arXiv.2305.13035},
	urldate = {2024-02-08},
	publisher = {arXiv},
	author = {Alabdulmohsin, Ibrahim and Zhai, Xiaohua and Kolesnikov, Alexander and Beyer, Lucas},
	month = jan,
	year = {2024},
}

@misc{zhai_scaling_2022,
	title = {Scaling {Vision} {Transformers}},
	url = {http://arxiv.org/abs/2106.04560},
	doi = {10.48550/arXiv.2106.04560},
	urldate = {2024-03-06},
	publisher = {arXiv},
	author = {Zhai, Xiaohua and Kolesnikov, Alexander and Houlsby, Neil and Beyer, Lucas},
	month = jun,
	year = {2022},
}

@inproceedings{raghu_transfusion_2019,
	title = {Transfusion: {Understanding} {Transfer} {Learning} for {Medical} {Imaging}},
	volume = {32},
	shorttitle = {Transfusion},
	url = {https://proceedings.neurips.cc/paper_files/paper/2019/hash/eb1e78328c46506b46a4ac4a1e378b91-Abstract.html},
	urldate = {2024-03-06},
	booktitle = {Advances in {Neural} {Information} {Processing} {Systems}},
	publisher = {Curran Associates, Inc.},
	author = {Raghu, Maithra and Zhang, Chiyuan and Kleinberg, Jon and Bengio, Samy},
	year = {2019},
}

@inproceedings{Masters2019a,
	title = {Twelve years of {Galaxy} {Zoo}},
	volume = {14},
	url = {http://arxiv.org/abs/1910.08177},
	doi = {10.1017/S1743921319008615},
	urldate = {2019-10-21},
	booktitle = {Proceedings of the {International} {Astronomical} {Union}},
	author = {Masters, Karen L.},
	month = oct,
	year = {2019},
	pages = {205--212},
}

@inproceedings{deng_imagenet_2009,
	title = {{ImageNet}: {A} large-scale hierarchical image database},
	shorttitle = {{ImageNet}},
	url = {https://ieeexplore.ieee.org/document/5206848},
	doi = {10.1109/CVPR.2009.5206848},
	urldate = {2024-03-07},
	booktitle = {2009 {IEEE} {Conference} on {Computer} {Vision} and {Pattern} {Recognition}},
	author = {Deng, Jia and Dong, Wei and Socher, Richard and Li, Li-Jia and Li, Kai and Fei-Fei, Li},
	month = jun,
	year = {2009},
	note = {ISSN: 1063-6919},
	pages = {248--255},
}

@article{andrew_galaxy_2023,
	title = {Galaxy {Classification} {Using} {Transfer} {Learning} and {Ensemble} of {CNNs} {With} {Multiple} {Colour} {Spaces}},
	copyright = {Creative Commons Attribution Non Commercial No Derivatives 4.0 International},
	url = {https://arxiv.org/abs/2305.00002},
	doi = {10.48550/ARXIV.2305.00002},
	urldate = {2024-01-15},
	author = {Andrew, Yevonnael},
	year = {2023},
}

@article{Dieleman2015,
	title = {Rotation-invariant convolutional neural networks for galaxy morphology prediction},
	volume = {450},
	issn = {0035-8711},
	url = {http://arxiv.org/abs/1503.07077},
	doi = {10.1093/mnras/stv632},
	number = {2},
	journal = {Monthly Notices of the Royal Astronomical Society},
	author = {Dieleman, S. and Willett, K. W. and Dambre, J.},
	year = {2015},
	pages = {1441--1459},
}

@article{Walmsley2022decals,
	title = {Galaxy {Zoo} {DECaLS}: {Detailed} visual morphology measurements from volunteers and deep learning for 314 000 galaxies},
	volume = {509},
	issn = {13652966},
	url = {https://doi.org/10.1093/mnras/stab2093},
	doi = {10.1093/mnras/stab2093},
	number = {3},
	urldate = {2021-02-16},
	journal = {Monthly Notices of the Royal Astronomical Society},
	author = {Walmsley, Mike and Lintott, Chris and Géron, Tobias and Kruk, Sandor and Krawczyk, Coleman and Willett, Kyle W. and Bamford, Steven and Kelvin, Lee S. and Fortson, Lucy and Gal, Yarin and Keel, William and Masters, Karen L. and Mehta, Vihang and Simmons, Brooke D. and Smethurst, Rebecca and Smith, Lewis and Baeten, Elisabeth M. and MacMillan, Christine},
	month = dec,
	year = {2022},
	pages = {3966--3988},
}

@article{Walmsley2023desi,
	title = {Galaxy {Zoo} {DESI}: {Detailed} {Morphology} {Measurements} for 8.{7M} {Galaxies} in the {DESI} {Legacy} {Imaging} {Surveys}},
	volume = {000},
	url = {http://arxiv.org/abs/2309.11425},
	urldate = {2023-09-29},
	journal = {MNRAS},
	author = {Walmsley, Mike and Géron, Tobias and Kruk, Sandor and Scaife, Anna M. M. and Lintott, Chris and Masters, Karen L. and Dawson, James M. and Dickinson, Hugh and Fortson, Lucy and Garland, Izzy L. and Mantha, Kameswara and O'Ryan, David and Popp, Jürgen and Simmons, Brooke and Baeten, Elisabeth M. and Macmillan, Christine},
	month = sep,
	year = {2023},
	pages = {1--18},
}

@inproceedings{He2016a,
	title = {Deep residual learning for image recognition},
	volume = {2016-Decem},
	isbn = {978-1-4673-8850-4},
	url = {http://ieeexplore.ieee.org/document/7780459/},
	doi = {10.1109/CVPR.2016.90},
	urldate = {2018-11-05},
	booktitle = {Proceedings of the {IEEE} {Computer} {Society} {Conference} on {Computer} {Vision} and {Pattern} {Recognition}},
	publisher = {IEEE},
	author = {He, Kaiming and Zhang, Xiangyu and Ren, Shaoqing and Sun, Jian},
	month = jun,
	year = {2016},
	pages = {770--778},
}

@article{Tan2021,
	title = {{EfficientNetV2}: {Smaller} {Models} and {Faster} {Training}},
	url = {http://arxiv.org/abs/2104.00298},
	urldate = {2022-01-13},
	author = {Tan, Mingxing and Le, Quoc V.},
	month = apr,
	year = {2021},
}

@article{Liu2022,
	title = {A {ConvNet} for the 2020s},
	url = {http://arxiv.org/abs/2201.03545},
	urldate = {2022-01-13},
	author = {Liu, Zhuang and Mao, Hanzi and Wu, Chao-Yuan and Feichtenhofer, Christoph and Darrell, Trevor and Xie, Saining},
	month = jan,
	year = {2022},
}

@inproceedings{Tu2022,
	title = {{MaxViT}: {Multi}-axis {Vision} {Transformer}},
	volume = {13684 LNCS},
	isbn = {978-3-031-20052-6},
	url = {https://arxiv.org/abs/2204.01697v4},
	doi = {10.1007/978-3-031-20053-3_27},
	urldate = {2022-10-13},
	booktitle = {Lecture {Notes} in {Computer} {Science} (including subseries {Lecture} {Notes} in {Artificial} {Intelligence} and {Lecture} {Notes} in {Bioinformatics})},
	author = {Tu, Zhengzhong and Talebi, Hossein and Zhang, Han and Yang, Feng and Milanfar, Peyman and Bovik, Alan and Li, Yinxiao},
	month = apr,
	year = {2022},
	pages = {459--479},
}

@article{Simmons2017,
	title = {Galaxy {Zoo}: {Quantitative} visual morphological classifications for 48 000 galaxies from {CANDELS}},
	volume = {464},
	issn = {13652966},
	doi = {10.1093/mnras/stw2587},
	number = {4},
	journal = {Monthly Notices of the Royal Astronomical Society},
	author = {Simmons, B. D. and Lintott, Chris and Willett, Kyle W. and Masters, Karen L. and Kartaltepe, Jeyhan S. and Häußler, Boris and Kaviraj, Sugata and Krawczyk, Coleman and Kruk, S. J. and McIntosh, Daniel H. and Smethurst, R. J. and Nichol, Robert C. and Scarlata, Claudia and Schawinski, Kevin and Conselice, Christopher J. and Almaini, Omar and Ferguson, Henry C. and Fortson, Lucy and Hartley, William and Kocevski, Dale and Koekemoer, Anton M. and Mortlock, Alice and Newman, Jeffrey A. and Bamford, Steven P. and Grogin, N. A. and Lucas, Ray A. and Hathi, Nimish P. and McGrath, Elizabeth and Peth, Michael and Pforr, Janine and Rizer, Zachary and Wuyts, Stijn and Barro, Guillermo and Bell, Eric F. and Castellano, Marco and Dahlen, Tomas and Dekel, Avishai and Ownsworth, Jamie and Faber, Sandra M. and Finkelstein, Steven L. and Fontana, Adriano and Galametz, Audrey and Grützbauch, Ruth and Koo, David and Lotz, Jennifer and Mobasher, Bahram and Mozena, Mark and Salvato, Mara and Wiklind, Tommy},
	year = {2017},
	pages = {4420--4447},
}

@article{Willett2017a,
	title = {Galaxy {Zoo}: {Morphological} classifications for 120 000 galaxies in {HST} legacy imaging},
	volume = {464},
	issn = {13652966},
	url = {https://academic.oup.com/mnras/article-lookup/doi/10.1093/mnras/stw2568},
	doi = {10.1093/mnras/stw2568},
	number = {4},
	urldate = {2019-01-22},
	journal = {Monthly Notices of the Royal Astronomical Society},
	author = {Willett, Kyle W. and Galloway, Melanie A. and Bamford, Steven P. and Lintott, Chris J. and Masters, Karen L. and Scarlata, Claudia and Simmons, B. D. and Beck, Melanie and Cardamone, Carolin N. and Cheung, Edmond and Edmondson, Edward M. and Fortson, Lucy F. and Griffith, Roger L. and Häußler, Boris and Han, Anna and Hart, Ross and Melvin, Thomas and Parrish, Michael and Schawinski, Kevin and Smethurst, R. J. and Smith, Arfon M.},
	month = feb,
	year = {2017},
	pages = {4176--4203},
}

@article{Willett2013,
	title = {Galaxy {Zoo} 2: {Detailed} morphological classifications for 304 122 galaxies from the sloan digital sky survey},
	volume = {435},
	issn = {00358711},
	doi = {10.1093/mnras/stt1458},
	number = {4},
	journal = {Monthly Notices of the Royal Astronomical Society},
	author = {Willett, Kyle W. and Lintott, Chris J. and Bamford, Steven P. and Masters, Karen L. and Simmons, Brooke D. and Casteels, Kevin R V and Edmondson, Edward M. and Fortson, Lucy F. and Kaviraj, Sugata and Keel, William C. and Melvin, Thomas and Nichol, Robert C. and Jordan Raddick, M. and Schawinski, Kevin and Simpson, Robert J. and Skibba, Ramin A. and Smith, Arfon M. and Thomas, Daniel},
	year = {2013},
	pages = {2835--2860},
}

@article{Walmsley2023zoobot,
	title = {Zoobot: {Adaptable} {Deep} {Learning} {Models} for {Galaxy} {Morphology}},
	volume = {8},
	url = {https://doi.org/10.21105/joss.05312},
	doi = {10.21105/joss.05312},
	number = {85},
	journal = {Journal of Open Source Software},
	author = {Walmsley, Mike and Allen, Campbell and Aussel, Ben and Bowles, Micah and Gregorowicz, Kasia and Slijepcevic, Inigo Val and Lintott, Chris J. and Scaife, Anna M. M. and Jabłońska, Maja and Karchev, Kosio and Lanzieri, Denise and Mohan, Devina and O’Ryan, David and Saiguhan, Bharath and Suárez, Crisel and Guerra-Varas, Nicolás and Velu, Renuka},
	year = {2023},
	note = {Publisher: The Open Journal},
	pages = {5312},
}

@article{Pawlik2016,
	title = {Shape asymmetry: {A} morphological indicator for automatic detection of galaxies in the post-coalescence merger stages},
	volume = {456},
	issn = {13652966},
	doi = {10.1093/mnras/stv2878},
	number = {3},
	journal = {Monthly Notices of the Royal Astronomical Society},
	author = {Pawlik, M. M. and Wild, V. and Walcher, C. J. and Johansson, P. H. and Villforth, C. and Rowlands, K. and Mendez-Abreu, J. and Hewlett, T.},
	year = {2016},
	pages = {3032--3052},
}

@article{Bottrell2019,
	title = {Deep learning predictions of galaxy merger stage and the importance of observational realism},
	volume = {490},
	issn = {13652966},
	url = {http://arxiv.org/abs/1910.07031},
	doi = {10.1093/mnras/stz2934},
	number = {4},
	urldate = {2019-10-17},
	journal = {Monthly Notices of the Royal Astronomical Society},
	author = {Bottrell, Connor and Hani, Maan H. and Teimoorinia, Hossen and Ellison, Sara L. and Moreno, Jorge and Torrey, Paul and Hayward, Christopher C. and Thorp, Mallory and Simard, Luc and Hernquist, Lars},
	month = oct,
	year = {2019},
	pages = {5390--5413},
}

@article{Pearson2019Mergers,
	title = {Identifying {Galaxy} {Mergers} in {Observations} and {Simulations} with {Deep} {Learning}},
	volume = {626},
	issn = {0004-6361},
	doi = {10.1051/0004-6361/201935355},
	number = {49},
	urldate = {2021-01-02},
	journal = {Astronomy \& Astrophysics},
	author = {Pearson, W. J. and Wang, L. and Trayford, J. W. and Petrillo, C. E. and van der Tak, F. F.S.},
	month = feb,
	year = {2019},
}

@article{Dey2018,
	title = {Overview of the {DESI} {Legacy} {Imaging} {Surveys}},
	volume = {157},
	issn = {00046256},
	url = {http://arxiv.org/abs/1804.08657},
	doi = {10.3847/1538-3881/ab089d},
	number = {5},
	urldate = {2018-09-24},
	journal = {The Astronomical Journal},
	author = {Dey, Arjun and Schlegel, David J. and Lang, Dustin and Blum, Robert and Burleigh, Kaylan and Fan, Xiaohui and Findlay, Joseph R. and Finkbeiner, Doug and Herrera, David and Juneau, Stéphanie and Landriau, Martin and Levi, Michael and McGreer, Ian and Meisner, Aaron and Myers, Adam D. and Moustakas, John and Nugent, Peter and Patej, Anna and Schlafly, Edward F. and Walker, Alistair R. and Valdes, Francisco and Weaver, Benjamin A. and Yèche, Christophe and Zou, Hu and Zhou, Xu and Abareshi, Behzad and Abbott, T. M. C. and Abolfathi, Bela and Aguilera, C. and Alam, Shadab and Allen, Lori and Alvarez, A. and Annis, James and Ansarinejad, Behzad and Aubert, Marie and Beechert, Jacqueline and Bell, Eric F. and BenZvi, Segev Y. and Beutler, Florian and Bielby, Richard M. and Bolton, Adam S. and Briceño, César and Buckley-Geer, Elizabeth J. and Butler, Karen and Calamida, Annalisa and Carlberg, Raymond G. and Carter, Paul and Casas, Ricard and Castander, Francisco J. and Choi, Yumi and Comparat, Johan and Cukanovaite, Elena and Delubac, Timothée and DeVries, Kaitlin and Dey, Sharmila and Dhungana, Govinda and Dickinson, Mark and Ding, Zhejie and Donaldson, John B. and Duan, Yutong and Duckworth, Christopher J. and Eftekharzadeh, Sarah and Eisenstein, Daniel J. and Etourneau, Thomas and Fagrelius, Parker A. and Farihi, Jay and Fitzpatrick, Mike and Font-Ribera, Andreu and Fulmer, Leah and Gänsicke, Boris T. and Gaztanaga, Enrique and George, Koshy and Gerdes, David W. and A Gontcho, Satya Gontcho and Gorgoni, Claudio and Green, Gregory and Guy, Julien and Harmer, Diane and Hernandez, M. and Honscheid, Klaus and Huang, Lijuan (Wendy) and James, David J. and Jannuzi, Buell T. and Jiang, Linhua and Joyce, Richard and Karcher, Armin and Karkar, Sonia and Kehoe, Robert and Kneib, Jean-Paul and Kueter-Young, Andrea and Lan, Ting-Wen and Lauer, Tod R. and Guillou, Laurent Le and Van Suu, Auguste Le and Lee, Jae Hyeon and Lesser, Michael and Levasseur, Laurence Perreault and Li, Ting S. and Mann, Justin L. and Marshall, Robert and Martínez-Vázquez, C. E. and Martini, Paul and du Mas des Bourboux, Hélion and McManus, Sean and Meier, Tobias Gabriel and Ménard, Brice and Metcalfe, Nigel and Muñoz-Gutiérrez, Andrea and Najita, Joan and Napier, Kevin and Narayan, Gautham and Newman, Jeffrey A. and Nie, Jundan and Nord, Brian and Norman, Dara J. and Olsen, Knut A. G. and Paat, Anthony and Palanque-Delabrouille, Nathalie and Peng, Xiyan and Poppett, Claire L. and Poremba, Megan R. and Prakash, Abhishek and Rabinowitz, David and Raichoor, Anand and Rezaie, Mehdi and Robertson, A. N. and Roe, Natalie A. and Ross, Ashley J. and Ross, Nicholas P. and Rudnick, Gregory and Safonova, Sasha and Saha, Abhijit and Sánchez, F. Javier and Savary, Elodie and Schweiker, Heidi and Scott, Adam and Seo, Hee-Jong and Shan, Huanyuan and Silva, David R. and Slepian, Zachary and Soto, Christian and Sprayberry, David and Staten, Ryan and Stillman, Coley M. and Stupak, Robert J. and Summers, David L. and Tie, Suk Sien and Tirado, H. and Vargas-Magaña, Mariana and Vivas, A. Katherina and Wechsler, Risa H. and Williams, Doug and Yang, Jinyi and Yang, Qian and Yapici, Tolga and Zaritsky, Dennis and Zenteno, A. and Zhang, Kai and Zhang, Tianmeng and Zhou, Rongpu and Zhou, Zhimin},
	month = apr,
	year = {2019},
	pages = {168},
}

@misc{loshchilov_decoupled_2019,
	title = {Decoupled {Weight} {Decay} {Regularization}},
	url = {http://arxiv.org/abs/1711.05101},
	doi = {10.48550/arXiv.1711.05101},
	urldate = {2024-03-13},
	publisher = {arXiv},
	author = {Loshchilov, Ilya and Hutter, Frank},
	month = jan,
	year = {2019},
}

@article{Atkinson2013,
	title = {Faint tidal features in galaxies within the {Canada}-{France}-{Hawaii} telescope legacy survey wide fields},
	volume = {765},
	issn = {15384357},
	doi = {10.1088/0004-637X/765/1/28},
	number = {1},
	journal = {Astrophysical Journal},
	author = {Atkinson, Adam M. and Abraham, Roberto G. and Ferguson, Annette M N},
	year = {2013},
}

@article{Buta2017catalog,
	title = {Galactic rings revisited - {I}. {CVRHS} classifications of 3962 ringed galaxies from the {Galaxy} {Zoo} 2 {Database}},
	volume = {471},
	issn = {13652966},
	url = {http://s3.amazonaws.com/zoo2/filen.jpg,},
	doi = {10.1093/MNRAS/STX1829},
	number = {4},
	urldate = {2021-09-22},
	journal = {Monthly Notices of the Royal Astronomical Society},
	author = {Buta, Ronald J.},
	month = nov,
	year = {2017},
	pages = {4027--4046},
}

@article{Lawrence2007,
	title = {The {UKIRT} infrared deep sky survey ({UKIDSS})},
	volume = {379},
	issn = {13652966},
	url = {https://academic.oup.com/mnras/article-lookup/doi/10.1111/j.1365-2966.2007.12040.x},
	doi = {10.1111/j.1365-2966.2007.12040.x},
	number = {4},
	urldate = {2018-05-23},
	journal = {Monthly Notices of the Royal Astronomical Society},
	author = {Lawrence, A. and Warren, S. J. and Almaini, O. and Edge, A. C. and Hambly, N. C. and Jameson, R. F. and Lucas, P. and Casali, M. and Adamson, A. and Dye, S. and Emerson, J. P. and Foucaud, S. and Hewett, P. and Hirst, P. and Hodgkin, S. T. and Irwin, M. J. and Lodieu, N. and McMahon, R. G. and Simpson, C. and Smail, I. and Mortlock, D. and Folger, M.},
	month = aug,
	year = {2007},
	pages = {1599--1617},
}

@misc{pandya_e2_2023,
	title = {E(2) {Equivariant} {Neural} {Networks} for {Robust} {Galaxy} {Morphology} {Classification}},
	copyright = {Creative Commons Attribution 4.0 International},
	url = {https://arxiv.org/abs/2311.01500},
	doi = {10.48550/ARXIV.2311.01500},
	urldate = {2024-01-15},
	author = {Pandya, Sneh and Patel, Purvik and O, Franc and Blazek, Jonathan},
	year = {2023},
}

@misc{gordon_uncovering_2024,
	title = {Uncovering Tidal Treasures: Automated Classification of Faint Tidal Features in {DECaLS} Data},
	url = {http://arxiv.org/abs/2404.06487},
	doi = {10.48550/arXiv.2404.06487},
	shorttitle = {Uncovering Tidal Treasures},
	number = {{arXiv}:2404.06487},
	publisher = {{arXiv}},
	author = {Gordon, Alexander J. and Ferguson, Annette M. N. and Mann, Robert G.},
    year = 2024,
	urldate = {2024-04-19},
	date = {2024-04-09},
	eprinttype = {arxiv},
	eprint = {2404.06487 [astro-ph]},
}

@article{Pedregosa2012,
	title = {Scikit-learn: {Machine} {Learning} in {Python}},
	volume = {12},
	issn = {15324435},
	url = {http://dl.acm.org/citation.cfm?id=2078195},
	doi = {10.1007/s13398-014-0173-7.2},
	journal = {Journal of Machine Learning Research},
	author = {Pedregosa, Fabian and Varoquaux, Gaël and Gramfort, Alexandre and Michel, Vincent and Thirion, Bertrand and Grisel, Olivier and Blondel, Mathieu and Prettenhofer, Peter and Weiss, Ron and Dubourg, Vincent and Vanderplas, Jake and Passos, Alexandre and Cournapeau, David and Brucher, Matthieu and Perrot, Matthieu and Duchesnay, Édouard},
	year = {2012},
	pmid = {1000044560},
	pages = {2825--2830},
}

@article{Hunter2007,
	title = {Matplotlib: {A} {2D} graphics environment},
	volume = {9},
	issn = {15219615},
	url = {http://ieeexplore.ieee.org/document/4160265/},
	doi = {10.1109/MCSE.2007.55},
	number = {3},
	urldate = {2018-11-01},
	journal = {Computing in Science and Engineering},
	author = {Hunter, John D.},
	month = may,
	year = {2007},
	pmid = {1000044628},
	pages = {90--95},
}

@article{VanderWalt2014,
	title = {Scikit-image: {Image} processing in python},
	volume = {2014},
	issn = {21678359},
	url = {https://peerj.com/articles/453},
	doi = {10.7717/peerj.453},
	number = {1},
	urldate = {2018-11-01},
	journal = {PeerJ},
	author = {Van Der Walt, Stéfan and Schönberger, Johannes L. and Nunez-Iglesias, Juan and Boulogne, François and Warner, Joshua D. and Yager, Neil and Gouillart, Emmanuelle and Yu, Tony},
	month = jun,
	year = {2014},
	pmid = {25024921},
	pages = {e453},
}

@misc{McKinney2010,
	title = {Data {Structures} for {Statistical} {Computing} in {Python}},
	urldate = {2018-11-01},
	author = {McKinney, Wes},
	year = {2010},
	doi = {10.25080/majora-92bf1922-00a},
	note = {Pages: 56-61
Publication Title: Proceedings of the 9th Python in Science Conference},
}

@inproceedings{Paszke2019,
	title = {{PyTorch}: {An} imperative style, high-performance deep learning library},
	volume = {32},
	url = {http://arxiv.org/abs/1912.01703},
	urldate = {2020-01-07},
	booktitle = {Advances in {Neural} {Information} {Processing} {Systems}},
	publisher = {Curran Associates, Inc.},
	author = {Paszke, Adam and Gross, Sam and Massa, Francisco and Lerer, Adam and Bradbury, James and Chanan, Gregory and Killeen, Trevor and Lin, Zeming and Gimelshein, Natalia and Antiga, Luca and Desmaison, Alban and Köpf, Andreas and Yang, Edward and DeVito, Zach and Raison, Martin and Tejani, Alykhan and Chilamkurthy, Sasank and Steiner, Benoit and Fang, Lu and Bai, Junjie and Chintala, Soumith},
	month = dec,
	year = {2019},
}

@article{bingham2018pyro,
	title = {Pyro: {Deep} {Universal} {Probabilistic} {Programming}},
	journal = {Journal of Machine Learning Research},
	author = {Bingham, Eli and Chen, Jonathan P and Jankowiak, Martin and Obermeyer, Fritz and Pradhan, Neeraj and Karaletsos, Theofanis and Singh, Rohit and Szerlip, Paul and Horsfall, Paul and Goodman, Noah D},
	year = {2018},
}

@article{TheAstropyCollaboration2018,
	title = {The {Astropy} {Project}: {Building} an inclusive, open-science project and status of the v2.0 core package},
	volume = {156},
	issn = {1538-3881},
	url = {http://arxiv.org/abs/1801.02634},
	doi = {arXiv:1801.02634v2},
	number = {3},
	urldate = {2018-09-14},
	journal = {The Astronomical Journal},
	author = {{The Astropy Collaboration} and Price-Whelan, A. M. and Sipőcz, B. M. and Günther, H. M. and Lim, P. L. and Crawford, S. M. and Conseil, S. and Shupe, D. L. and Craig, M. W. and Dencheva, N. and Ginsburg, A. and VanderPlas, J. T. and Bradley, L. D. and Pérez-Suárez, D. and de Val-Borro, M. and Aldcroft, T. L. and Cruz, K. L. and Robitaille, T. P. and Tollerud, E. J. and Ardelean, C. and Babej, T. and Bachetti, M. and Bakanov, A. V. and Bamford, S. P. and Barentsen, G. and Barmby, P. and Baumbach, A. and Berry, K. L. and Biscani, F. and Boquien, M. and Bostroem, K. A. and Bouma, L. G. and Brammer, G. B. and Bray, E. M. and Breytenbach, H. and Buddelmeijer, H. and Burke, D. J. and Calderone, G. and Rodríguez, J. L. Cano and Cara, M. and Cardoso, J. V. M. and Cheedella, S. and Copin, Y. and Crichton, D. and DÁvella, D. and Deil, C. and Depagne, É. and Dietrich, J. P. and Donath, A. and Droettboom, M. and Earl, N. and Erben, T. and Fabbro, S. and Ferreira, L. A. and Finethy, T. and Fox, R. T. and Garrison, L. H. and Gibbons, S. L. J. and Goldstein, D. A. and Gommers, R. and Greco, J. P. and Greenfield, P. and Groener, A. M. and Grollier, F. and Hagen, A. and Hirst, P. and Homeier, D. and Horton, A. J. and Hosseinzadeh, G. and Hu, L. and Hunkeler, J. S. and Ivezić, Ž. and Jain, A. and Jenness, T. and Kanarek, G. and Kendrew, S. and Kern, N. S. and Kerzendorf, W. E. and Khvalko, A. and King, J. and Kirkby, D. and Kulkarni, A. M. and Kumar, A. and Lee, A. and Lenz, D. and Littlefair, S. P. and Ma, Z. and Macleod, D. M. and Mastropietro, M. and McCully, C. and Montagnac, S. and Morris, B. M. and Mueller, M. and Mumford, S. J. and Muna, D. and Murphy, N. A. and Nelson, S. and Nguyen, G. H. and Ninan, J. P. and Nöthe, M. and Ogaz, S. and Oh, S. and Parejko, J. K. and Parley, N. and Pascual, S. and Patil, R. and Patil, A. A. and Plunkett, A. L. and Prochaska, J. X. and Rastogi, T. and Janga, V. Reddy and Sabater, J. and Sakurikar, P. and Seifert, M. and Sherbert, L. E. and Sherwood-Taylor, H. and Shih, A. Y. and Sick, J. and Silbiger, M. T. and Singanamalla, S. and Singer, L. P. and Sladen, P. H. and Sooley, K. A. and Sornarajah, S. and Streicher, O. and Teuben, P. and Thomas, S. W. and Tremblay, G. R. and Turner, J. E. H. and Terrón, V. and van Kerkwijk, M. H. and de la Vega, A. and Watkins, L. L. and Weaver, B. A. and Whitmore, J. B. and Woillez, J. and Zabalza, V.},
	month = aug,
	year = {2018},
	pages = {123},
}

@article{virtanen_scipy_2020,
	title = {{SciPy} 1.0: {Fundamental} {Algorithms} for {Scientific} {Computing} in {Python}},
	volume = {17},
	doi = {10.1038/s41592-019-0686-2},
	journal = {Nature Methods},
	author = {Virtanen, Pauli and Gommers, Ralf and Oliphant, Travis E. and Haberland, Matt and Reddy, Tyler and Cournapeau, David and Burovski, Evgeni and Peterson, Pearu and Weckesser, Warren and Bright, Jonathan and van der Walt, Stéfan J. and Brett, Matthew and Wilson, Joshua and Millman, K. Jarrod and Mayorov, Nikolay and Nelson, Andrew R. J. and Jones, Eric and Kern, Robert and Larson, Eric and Carey, C J and Polat, İlhan and Feng, Yu and Moore, Eric W. and VanderPlas, Jake and Laxalde, Denis and Perktold, Josef and Cimrman, Robert and Henriksen, Ian and Quintero, E. A. and Harris, Charles R. and Archibald, Anne M. and Ribeiro, Antônio H. and Pedregosa, Fabian and van Mulbregt, Paul and {SciPy 1.0 Contributors}},
	year = {2020},
	pages = {261--272},
}

@article{bressem_comparing_2020,
	title = {Comparing different deep learning architectures for classification of chest radiographs},
	volume = {10},
	copyright = {2020 The Author(s)},
	issn = {2045-2322},
	url = {https://www.nature.com/articles/s41598-020-70479-z},
	doi = {10.1038/s41598-020-70479-z},
	language = {en},
	number = {1},
	urldate = {2024-06-01},
	journal = {Scientific Reports},
	author = {Bressem, Keno K. and Adams, Lisa C. and Erxleben, Christoph and Hamm, Bernd and Niehues, Stefan M. and Vahldiek, Janis L.},
	month = aug,
	year = {2020},
	note = {Publisher: Nature Publishing Group},
	pages = {13590},
}

@inproceedings{ke_chextransfer_2021,
	address = {New York, NY, USA},
	series = {{CHIL} '21},
	title = {{CheXtransfer}: performance and parameter efficiency of {ImageNet} models for chest {X}-{Ray} interpretation},
	isbn = {978-1-4503-8359-2},
	shorttitle = {{CheXtransfer}},
	url = {https://dl.acm.org/doi/10.1145/3450439.3451867},
	doi = {10.1145/3450439.3451867},
	urldate = {2024-06-01},
	booktitle = {Proceedings of the {Conference} on {Health}, {Inference}, and {Learning}},
	publisher = {Association for Computing Machinery},
	author = {Ke, Alexander and Ellsworth, William and Banerjee, Oishi and Ng, Andrew Y. and Rajpurkar, Pranav},
	month = apr,
	year = {2021},
	pages = {116--124},
}

@article{fang_does_2023,
	title = {Does progress on {ImageNet} transfer to real-world datasets?},
	volume = {36},
	url = {https://proceedings.neurips.cc/paper_files/paper/2023/hash/4eb33c53ed5b14ce9028309431f565cc-Abstract-Datasets_and_Benchmarks.html},
	language = {en},
	urldate = {2024-05-22},
	journal = {Advances in Neural Information Processing Systems},
	author = {Fang, Alex and Kornblith, Simon and Schmidt, Ludwig},
	month = dec,
	year = {2023},
	pages = {25050--25080},
}

@article{tuggener_is_2022,
	title = {Is it enough to optimize {CNN} architectures on {ImageNet}?},
	volume = {4},
	issn = {2624-9898},
	url = {http://arxiv.org/abs/2103.09108},
	doi = {10.3389/fcomp.2022.1041703},
	urldate = {2024-06-01},
	journal = {Frontiers in Computer Science},
	author = {Tuggener, Lukas and Schmidhuber, Jürgen and Stadelmann, Thilo},
	month = nov,
	year = {2022},
	pages = {1041703},
}

@article{HuertasCompany2022,
	title = {The {Dawes} {Review} 10: {The} impact of deep learning for the analysis of galaxy surveys},
	volume = {40},
	issn = {14486083},
	url = {http://arxiv.org/abs/2210.01813},
	doi = {10.1017/pasa.2022.55},
	urldate = {2022-10-07},
	journal = {Publications of the Astronomical Society of Australia},
	author = {Huertas-Company, M. and Lanusse, F.},
	month = oct,
	year = {2023},
}

@misc{walmsley_scaling_2024,
	title = {Scaling {Laws} for {Galaxy} {Images}},
	url = {https://arxiv.org/abs/2404.02973v1},
	language = {en},
	urldate = {2024-06-05},
	journal = {arXiv.org},
	author = {Walmsley, Mike and Bowles, Micah and Scaife, Anna M. M. and Makechemu, Jason Shingirai and Gordon, Alexander J. and Ferguson, Annette M. N. and Mann, Robert G. and Pearson, James and Popp, Jürgen J. and Bovy, Jo and Speagle, Josh and Dickinson, Hugh and Fortson, Lucy and Géron, Tobias and Kruk, Sandor and Lintott, Chris J. and Mantha, Kameswara and Mohan, Devina and O'Ryan, David and Slijepevic, Inigo V.},
	month = apr,
	year = {2024},
}

@incollection{Buta2013,
	title = {Galaxy morphology},
	isbn = {978-94-007-5609-0},
	url = {https://link.springer.com/referenceworkentry/10.1007/978-94-007-5609-0_1},
	urldate = {2021-09-22},
	booktitle = {Planets, {Stars} and {Stellar} {Systems}: {Volume} 6: {Extragalactic} {Astronomy} and {Cosmology}},
	publisher = {Springer Netherlands},
	author = {Buta, Ronald J.},
	month = jan,
	year = {2013},
	doi = {10.1007/978-94-007-5609-0_1},
	pages = {1--89},
}

@article{li_galaxy_2023,
	title = {Galaxy morphology classification using multiscale convolution capsule network},
	volume = {523},
	issn = {0035-8711, 1365-2966},
	url = {https://academic.oup.com/mnras/article/523/1/488/7179754},
	doi = {10.1093/mnras/stad854},
	language = {en},
	number = {1},
	urldate = {2024-01-15},
	journal = {Monthly Notices of the Royal Astronomical Society},
	author = {Li, Guangping and Xu, Tingting and Li, Liping and Gao, Xianjun and Liu, Zhijing and Cao, Jie and Yang, Mingcun and Zhou, Weihong},
	month = may,
	year = {2023},
	pages = {488--497},
}

@article{Katebi2019a,
	title = {Galaxy morphology prediction using {Capsule} {Networks}},
	volume = {486},
	issn = {13652966},
	url = {https://academic.oup.com/mnras/article/486/2/1539/5424782},
	doi = {10.1093/mnras/stz915},
	number = {2},
	urldate = {2019-05-17},
	journal = {Monthly Notices of the Royal Astronomical Society},
	author = {Katebi, Reza and Zhou, Yadi and Chornock, Ryan and Bunescu, Razvan},
	month = jun,
	year = {2019},
	pages = {1539--1547},
}

@misc{houlsby_bayesian_2011,
	title = {Bayesian {Active} {Learning} for {Classification} and {Preference} {Learning}},
	url = {https://arxiv.org/abs/1112.5745v1},
	language = {en},
	urldate = {2024-06-05},
	journal = {arXiv.org},
	author = {Houlsby, Neil and Huszár, Ferenc and Ghahramani, Zoubin and Lengyel, Máté},
	month = dec,
	year = {2011},
}

@article{Abazajian2009,
	title = {the {Seventh} {Data} {Release} of the {Sloan} {Digital} {Sky} {Survey}},
	volume = {182},
	issn = {0067-0049},
	url = {http://stacks.iop.org/0067-0049/182/i=2/a=543?key=crossref.bc07496fb06b943bcf82755687fa84b4},
	doi = {10.1088/0067-0049/182/2/543},
	number = {2},
	journal = {The Astrophysical Journal Supplement Series},
	author = {Abazajian, Kevork N. and Adelman-McCarthy, Jennifer K. and Agüeros, Marcel A. and Allam, Sahar S. and Prieto, Carlos Allende and An, Deokkeun and Anderson, Kurt S. J. and Anderson, Scott F. and Annis, James and Bahcall, Neta A. and Bailer-Jones, C. A. L. and Barentine, J. C. and Bassett, Bruce A. and Becker, Andrew C. and Beers, Timothy C. and Bell, Eric F. and Belokurov, Vasily and Berlind, Andreas A. and Berman, Eileen F. and Bernardi, Mariangela and Bickerton, Steven J. and Bizyaev, Dmitry and Blakeslee, John P. and Blanton, Michael R. and Bochanski, John J. and Boroski, William N. and Brewington, Howard J. and Brinchmann, Jarle and Brinkmann, J. and Brunner, Robert J. and Budavári, Tamás and Carey, Larry N. and Carliles, Samuel and Carr, Michael A. and Castander, Francisco J. and Cinabro, David and Connolly, A. J. and Csabai, István and Cunha, Carlos E. and Czarapata, Paul C. and Davenport, James R. A. and de Haas, Ernst and Dilday, Ben and Doi, Mamoru and Eisenstein, Daniel J. and Evans, Michael L. and Evans, N. W. and Fan, Xiaohui and Friedman, Scott D. and Frieman, Joshua A. and Fukugita, Masataka and Gänsicke, Boris T. and Gates, Evalyn and Gillespie, Bruce and Gilmore, G. and Gonzalez, Belinda and Gonzalez, Carlos F. and Grebel, Eva K. and Gunn, James E. and Györy, Zsuzsanna and Hall, Patrick B. and Harding, Paul and Harris, Frederick H. and Harvanek, Michael and Hawley, Suzanne L. and Hayes, Jeffrey J. E. and Heckman, Timothy M. and Hendry, John S. and Hennessy, Gregory S. and Hindsley, Robert B. and Hoblitt, J. and Hogan, Craig J. and Hogg, David W. and Holtzman, Jon A. and Hyde, Joseph B. and Ichikawa, Shin-ichi and Ichikawa, Takashi and Im, Myungshin and Ivezic, Zeljko and Jester, Sebastian and Jiang, Linhua and Johnson, Jennifer A. and Jorgensen, Anders M. and Jurić, Mario and Kent, Stephen M. and Kessler, R. and Kleinman, S. J. and Knapp, G. R. and Konishi, Kohki and Kron, Richard G. and Krzesinski, Jurek and Kuropatkin, Nikolay and Lampeitl, Hubert and Lebedeva, Svetlana and Lee, Myung Gyoon and Lee, Young Sun and Leger, R. French and Lépine, Sébastien and Li, Nolan and Lima, Marcos and Lin, Huan and Long, Daniel C. and Loomis, Craig P. and Loveday, Jon and Lupton, Robert H. and Magnier, Eugene and Malanushenko, Olena and Malanushenko, Viktor and Mandelbaum, Rachel and Margon, Bruce and Marriner, John P. and Martínez-Delgado, David and Matsubara, Takahiko and McGehee, Peregrine M. and McKay, Timothy A. and Meiksin, Avery and Morrison, Heather L. and Mullally, Fergal and Munn, Jeffrey A. and Murphy, Tara and Nash, Thomas and Nebot, Ada and Neilsen, Eric H. and Newberg, Heidi Jo and Newman, Peter R. and Nichol, Robert C. and Nicinski, Tom and Nieto-Santisteban, Maria and Nitta, Atsuko and Okamura, Sadanori and Oravetz, Daniel J. and Ostriker, Jeremiah P. and Owen, Russell and Padmanabhan, Nikhil and Pan, Kaike and Park, Changbom and Pauls, George and Peoples, John and Percival, Will J. and Pier, Jeffrey R. and Pope, Adrian C. and Pourbaix, Dimitri and Price, Paul A. and Purger, Norbert and Quinn, Thomas and Raddick, M. Jordan and Fiorentin, Paola Re and Richards, Gordon T. and Richmond, Michael W. and Riess, Adam G. and Rix, Hans-Walter and Rockosi, Constance M. and Sako, Masao and Schlegel, David J. and Schneider, Donald P. and Scholz, Ralf-Dieter and Schreiber, Matthias R. and Schwope, Axel D. and Seljak, Uroš and Sesar, Branimir and Sheldon, Erin and Shimasaku, Kazu and Sibley, Valena C. and Simmons, A. E. and Sivarani, Thirupathi and Smith, J. Allyn and Smith, Martin C. and Smolčić, Vernesa and Snedden, Stephanie A. and Stebbins, Albert and Steinmetz, Matthias and Stoughton, Chris and Strauss, Michael A. and SubbaRao, Mark and Suto, Yasushi and Szalay, Alexander S. and Szapudi, István and Szkody, Paula and Tanaka, Masayuki and Tegmark, Max and Teodoro, Luis F. A. and Thakar, Aniruddha R. and Tremonti, Christy A. and Tucker, Douglas L. and Uomoto, Alan and Vanden Berk, Daniel E. and Vandenberg, Jan and Vidrih, S. and Vogeley, Michael S. and Voges, Wolfgang and Vogt, Nicole P. and Wadadekar, Yogesh and Watters, Shannon and Weinberg, David H. and West, Andrew A. and White, Simon D. M. and Wilhite, Brian C. and Wonders, Alainna C. and Yanny, Brian and Yocum, D. R. and York, Donald G. and Zehavi, Idit and Zibetti, Stefano and Zucker, Daniel B.},
	year = {2009},
	pages = {543--558},
}

@article{walmsley_galaxy_2020,
	title = {Galaxy {Zoo}: {Probabilistic} {Morphology} through {Bayesian} {CNNs} and {Active} {Learning}},
	volume = {491},
	issn = {0035-8711, 1365-2966},
	shorttitle = {Galaxy {Zoo}},
	url = {http://arxiv.org/abs/1905.07424},
	doi = {10.1093/mnras/stz2816},
	number = {2},
	urldate = {2024-06-05},
	journal = {Monthly Notices of the Royal Astronomical Society},
	author = {Walmsley, Mike and Smith, Lewis and Lintott, Chris and Gal, Yarin and Bamford, Steven and Dickinson, Hugh and Fortson, Lucy and Kruk, Sandor and Masters, Karen and Scarlata, Claudia and Simmons, Brooke and Smethurst, Rebecca and Wright, Darryl},
	month = jan,
	year = {2020},
	pages = {1554--1574},
}

@misc{raddick_galaxy_2013,
	title = {Galaxy {Zoo}: {Motivations} of {Citizen} {Scientists}},
	shorttitle = {Galaxy {Zoo}},
	url = {http://arxiv.org/abs/1303.6886},
	doi = {10.48550/arXiv.1303.6886},
	urldate = {2024-06-04},
	publisher = {arXiv},
	author = {Raddick, M. Jordan and Bracey, Georgia and Gay, Pamela L. and Lintott, Chris J. and Cardamone, Carie and Murray, Phil and Schawinski, Kevin and Szalay, Alexander S. and Vandenberg, Jan},
	month = mar,
	year = {2013},
}

@misc{recht_imagenet_2019,
	title = {Do {ImageNet} {Classifiers} {Generalize} to {ImageNet}?},
	url = {http://arxiv.org/abs/1902.10811},
	doi = {10.48550/arXiv.1902.10811},
	urldate = {2024-03-15},
	publisher = {arXiv},
	author = {Recht, Benjamin and Roelofs, Rebecca and Schmidt, Ludwig and Shankar, Vaishaal},
	month = jun,
	year = {2019},
}

@article{Aitchison2021,
	title = {a {Statistical} {Theory} of {Cold} {Posteriors} in {Deep} {Neural} {Networks}},
	url = {https://openreview.net/forum?id=Rd138pWXMvG},
	urldate = {2022-01-19},
	journal = {ICLR 2021 - 9th International Conference on Learning Representations},
	author = {Aitchison, Laurence},
	year = {2021},
}

@article{kasempakdeepong_sugarcane_2022,
	title = {Sugarcane {Classification} for {On}-{Site} {Assessment} {Using} {Computer} {Vision}},
	copyright = {https://doi.org/10.15223/policy-029},
	url = {https://ieeexplore.ieee.org/document/9960252/},
	doi = {10.1109/iSAI-NLP56921.2022.9960252},
	urldate = {2024-06-06},
	journal = {2022 17th International Joint Symposium on Artificial Intelligence and Natural Language Processing (iSAI-NLP)},
	author = {Kasempakdeepong, Piyapoj and Ponchaiyapruek, Pondsulee and Viriyothai, Pattamon and Songchumrong, Anuwat and Kantavat, Pittipol and Pungprasertying, Prasertsak},
	month = nov,
	year = {2022},
	note = {Conference Name: 2022 17th International Joint Symposium on Artificial Intelligence and Natural Language Processing (iSAI-NLP)
ISBN: 9781665457279
Place: Chiang Mai, Thailand
Publisher: IEEE},
	pages = {1--6},
}

@article{mezher_novel_2023,
	title = {A {Novel} {Strategy} for {Improving} {Robustness} in {Computer} {Vision} {Manufacturing} {Defect} {Detection}},
	copyright = {Creative Commons Attribution 4.0 International},
	url = {https://arxiv.org/abs/2305.09407},
	doi = {10.48550/ARXIV.2305.09407},
	urldate = {2024-06-06},
	author = {Mezher, Ahmad Mohamad and Marble, Andrew E.},
	year = {2023},
}

@misc{jackson_unleashing_2024,
	address = {Rochester, NY},
	type = {{SSRN} {Scholarly} {Paper}},
	title = {Unleashing the {Power} of the {Zooniverse}: {The} 2021 {Survey} of {Volunteers}},
	shorttitle = {Unleashing the {Power} of the {Zooniverse}},
	url = {https://papers.ssrn.com/abstract=4830179},
	doi = {10.2139/ssrn.4830179},
	language = {en},
	urldate = {2024-06-04},
	author = {Jackson, Corey and Dowthwaite, Liz and Jeong, Ellie and Trouille, Laura and Fortson, Lucy and Lintott, Chris and Simmons, Brooke and Miller, Grant},
	month = may,
	year = {2024},
}

@article{jeong_how_2024,
	title = {How {Personal} {Value} {Orientations} {Influence} {Behaviors} in {Digital} {Citizen} {Science}},
	volume = {8},
	url = {https://dl.acm.org/doi/10.1145/3637341},
	doi = {10.1145/3637341},
	number = {CSCW1},
	urldate = {2024-06-04},
	journal = {Proceedings of the ACM on Human-Computer Interaction},
	author = {Jeong, Eunmi (Ellie) and Jackson, Corey and Dowthwaite, Liz and Johnson, Cliff and Trouille, Laura},
	month = apr,
	year = {2024},
	pages = {64:1--64:25},
}

@misc{huang_deep_2016,
	title = {Deep {Networks} with {Stochastic} {Depth}},
	url = {http://arxiv.org/abs/1603.09382},
	doi = {10.48550/arXiv.1603.09382},
	urldate = {2024-02-12},
	publisher = {arXiv},
	author = {Huang, Gao and Sun, Yu and Liu, Zhuang and Sedra, Daniel and Weinberger, Kilian},
	month = jul,
	year = {2016},
}

@misc{gebru_datasheets_2021,
	title = {Datasheets for {Datasets}},
	url = {http://arxiv.org/abs/1803.09010},
	doi = {10.48550/arXiv.1803.09010},
	urldate = {2024-06-10},
	publisher = {arXiv},
	author = {Gebru, Timnit and Morgenstern, Jamie and Vecchione, Briana and Vaughan, Jennifer Wortman and Wallach, Hanna and Daumé III, Hal and Crawford, Kate},
	month = dec,
	year = {2021},
}

@article{Stein2022,
	title = {Mining for {Strong} {Gravitational} {Lenses} with {Self}-supervised {Learning}},
	volume = {932},
	issn = {0004-637X},
	url = {http://arxiv.org/abs/2110.00023},
	doi = {10.3847/1538-4357/ac6d63},
	number = {2},
	journal = {The Astrophysical Journal},
	author = {Stein, George and Blaum, Jacqueline and Harrington, Peter and Medan, Tomislav and Lukić, Zarija},
	year = {2022},
	pages = {107},
}

@misc{smith_astropt_2024,
	title = {{AstroPT}: {Scaling} {Large} {Observation} {Models} for {Astronomy}},
	shorttitle = {{AstroPT}},
	url = {http://arxiv.org/abs/2405.14930},
	urldate = {2024-05-27},
	publisher = {arXiv},
	author = {Smith, Michael J. and Roberts, Ryan J. and Angeloudi, Eirini and Huertas-Company, Marc},
	month = may,
	year = {2024},
}

@misc{euclid_collaboration_euclid_2025,
	       author = {{Euclid Collaboration: Siudek}, M. and {Huertas-Company}, M. and {Smith}, M. and others},
       title = "{Euclid Quick Data Release (Q1) Exploring galaxy properties with a multi-modal foundation model}",
      journal = {A\&A, in press (Euclid Q1 SI), \url{https://doi.org/10.1051/0004-6361/202554611}},
         year = 2025,
        month = mar,
          eid = {arXiv:2503.15312},
        pages = {arXiv:2503.15312},
archivePrefix = {arXiv},
       eprint = {2503.15312},
 primaryClass = {astro-ph.GA},
       adsurl = {https://ui.adsabs.harvard.edu/abs/2025arXiv250315312E}
}

@article{parker_astroclip_2024,
	title = {{AstroCLIP}: a cross-modal foundation model for galaxies},
	volume = {531},
	issn = {0035-8711},
	shorttitle = {{AstroCLIP}},
	url = {https://doi.org/10.1093/mnras/stae1450},
	doi = {10.1093/mnras/stae1450},
	number = {4},
	urldate = {2024-06-30},
	journal = {Monthly Notices of the Royal Astronomical Society},
	author = {Parker, Liam and Lanusse, Francois and Golkar, Siavash and Sarra, Leopoldo and Cranmer, Miles and Bietti, Alberto and Eickenberg, Michael and Krawezik, Geraud and McCabe, Michael and Morel, Rudy and Ohana, Ruben and Pettee, Mariel and Régaldo-Saint Blancard, Bruno and Cho, Kyunghyun and Ho, Shirley and {The Polymathic AI Collaboration}},
	month = jul,
	year = {2024},
	pages = {4990--5011},
}

@inproceedings{angeloudi_multimodal_2024,
	title = {The {Multimodal} {Universe}: {Enabling} {Large}-{Scale} {Machine} {Learning} with 100 {TB} of {Astronomical} {Scientific} {Data}},
	volume = {37},
	url = {https://proceedings.neurips.cc/paper_files/paper/2024/file/6a57493d35fefea59d06396c7cb69228-Paper-Datasets_and_Benchmarks_Track.pdf},
	booktitle = {Advances in {Neural} {Information} {Processing} {Systems}},
	publisher = {Curran Associates, Inc.},
	author = {Angeloudi, Eirini and Audenaert, Jeroen and Bowles, Micah and Boyd, Benjamin M. and Chemaly, David and Cherinka, Brian and Ciucă, Ioana and Cranmer, Miles and Do, Aaron and Grayling, Matthew and Hayes, Erin E. and Hehir, Tom and Ho, Shirley and Huertas-Company, Marc and Iyer, Kartheik G. and Jablonska, Maja and Lanusse, Francois and Leung, Henry W. and Mandel, Kaisey and Martínez-Galarza, Juan Rafael and Melchior, Peter and Meyer, Lucas and Parker, Liam H. and Qu, Helen and Shen, Jeff and Smith, Michael J. and Stone, Connor and Walmsley, Mike and Wu, John F.},
	editor = {Globerson, A. and Mackey, L. and Belgrave, D. and Fan, A. and Paquet, U. and Tomczak, J. and Zhang, C.},
	year = {2024},
	pages = {57841--57913},
}

@article{masters_galaxy_2024,
	title = {Galaxy {Zoo}: {Morphologies} based on {UKIDSS} {NIR} {Imaging} for 71,052 {Galaxies}},
	volume = {8},
	issn = {2515-5172},
	shorttitle = {Galaxy {Zoo}},
	url = {http://arxiv.org/abs/2408.10160},
	doi = {10.3847/2515-5172/ad6f10},
	number = {8},
	urldate = {2025-01-15},
	journal = {Research Notes of the AAS},
	author = {Masters, Karen L. and Galloway, Melanie and Fortson, Lucy and Lintott, Chris and Read, Mike and Scarlata, Claudia and Simmons, Brooke and Walmsley, Mike and Willett, Kyle},
	month = aug,
	year = {2024},
	pages = {198},
}

@misc{wightman_resnet_2021,
	title = {{ResNet} strikes back: {An} improved training procedure in timm},
	shorttitle = {{ResNet} strikes back},
	url = {http://arxiv.org/abs/2110.00476},
	doi = {10.48550/arXiv.2110.00476},
	urldate = {2024-03-31},
	publisher = {arXiv},
	author = {Wightman, Ross and Touvron, Hugo and Jégou, Hervé},
	month = oct,
	year = {2021},
}

@misc{chenPaLIJointlyScaledMultilingual2023,
	title = {{PaLI}: {A} {Jointly}-{Scaled} {Multilingual} {Language}-{Image} {Model}},
	shorttitle = {{PaLI}},
	url = {http://arxiv.org/abs/2209.06794},
	doi = {10.48550/arXiv.2209.06794},
	urldate = {2025-05-13},
	publisher = {arXiv},
	author = {Chen, Xi and Wang, Xiao and Changpinyo, Soravit and Piergiovanni, A. J. and Padlewski, Piotr and Salz, Daniel and Goodman, Sebastian and Grycner, Adam and Mustafa, Basil and Beyer, Lucas and Kolesnikov, Alexander and Puigcerver, Joan and Ding, Nan and Rong, Keran and Akbari, Hassan and Mishra, Gaurav and Xue, Linting and Thapliyal, Ashish and Bradbury, James and Kuo, Weicheng and Seyedhosseini, Mojtaba and Jia, Chao and Ayan, Burcu Karagol and Riquelme, Carlos and Steiner, Andreas and Angelova, Anelia and Zhai, Xiaohua and Houlsby, Neil and Soricut, Radu},
	month = jun,
	year = {2023},
}

@misc{tschannenSigLIP2Multilingual2025,
	title = {{SigLIP} 2: {Multilingual} {Vision}-{Language} {Encoders} with {Improved} {Semantic} {Understanding}, {Localization}, and {Dense} {Features}},
	shorttitle = {{SigLIP} 2},
	url = {http://arxiv.org/abs/2502.14786},
	doi = {10.48550/arXiv.2502.14786},
	urldate = {2025-05-13},
	publisher = {arXiv},
	author = {Tschannen, Michael and Gritsenko, Alexey and Wang, Xiao and Naeem, Muhammad Ferjad and Alabdulmohsin, Ibrahim and Parthasarathy, Nikhil and Evans, Talfan and Beyer, Lucas and Xia, Ye and Mustafa, Basil and Hénaff, Olivier and Harmsen, Jeremiah and Steiner, Andreas and Zhai, Xiaohua},
	month = feb,
	year = {2025},
}

@inproceedings{Agrawal2022VisionLanguagePC,
  title={Vision-Language Pretraining: Current Trends and the Future},
  author={Aishwarya Agrawal and Damien Teney and Aida Nematzadeh},
  booktitle={Annual Meeting of the Association for Computational Linguistics},
  year={2022},
  url={https://api.semanticscholar.org/CorpusID:248780422}
}

@article{Xiao2024ASIMSAAS,
  title={ASIMSA: Advanced Semantic Information Guided Multi-Scale Alignment Framework for Medical Vision-Language Pretraining},
  author={Shuai Xiao and Yangyang Zhang and Liming Jiang and Zhengxia Wang},
  journal={2024 IEEE 9th International Conference on Computational Intelligence and Applications (ICCIA)},
  year={2024},
  pages={99-103},
  url={https://api.semanticscholar.org/CorpusID:273534365}
}

@inproceedings{Yan2025MAKEMK,
  title={MAKE: Multi-Aspect Knowledge-Enhanced Vision-Language Pretraining for Zero-shot Dermatological Assessment},
  author={Siyuan Yan and Xieji Li and Ming Hu and Yiwen Jiang and Zhen Yu and Zongyuan Ge},
  year={2025},
  url={https://api.semanticscholar.org/CorpusID:278602553}
}

@article{Liu2023ImprovingMV,
  title={Improving Medical Vision-Language Contrastive Pretraining With Semantics-Aware Triage},
  author={Bo Liu and Donghuan Lu and Dong Wei and Xianming Wu and Yan Wang and Yu Zhang and Yefeng Zheng},
  journal={IEEE Transactions on Medical Imaging},
  year={2023},
  volume={42},
  pages={3579-3589},
  url={https://api.semanticscholar.org/CorpusID:259856982}
}

@article{Cho2024PretrainingVM,
  title={Pretraining Vision-Language Model for Difference Visual Question Answering in Longitudinal Chest X-rays},
  author={Yeongjae Cho and Taehee Kim and Heejun Shin and Sungzoon Cho and Dongmyung Shin},
  journal={ArXiv},
  year={2024},
  volume={abs/2402.08966},
  url={https://api.semanticscholar.org/CorpusID:267657514}
}

@article{AtFirstSight2024,
  title={At First Sight! Zero-shot Classification of Astronomical Images with Large Multimodal Models},
  author={Dimitrios Tanoglidis and Bhuvnesh Jain},
  journal={Research Notes of the AAS},
  year={2024},
  volume={8},
  issue={10},
  url={https://iopscience.iop.org/article/10.3847/2515-5172/ad887a}
}

@ARTICLE{Q1-SP048,
   author = {{Euclid Collaboration: Walmsley}, M. and {Holloway}, P. and {Lines}, N.~E.~P. and others},
   title = "{Euclid Quick Data Release (Q1): The Strong Lensing Discovery Engine A -- System overview and lens catalogue}",
  journal = {A\&A, accepted (Euclid Q1 SI)},
     year = 2025,
    month = mar,
      eid = {arXiv:2503.15324},
    pages = {arXiv:2503.15324},
archivePrefix = {arXiv},
   eprint = {2503.15324},
primaryClass = {astro-ph.GA},
   adsurl = {https://ui.adsabs.harvard.edu/abs/2025arXiv250315324E}
}

@ARTICLE{Q1-SP052,
       author = {{Euclid Collaboration: Rojas}, K. and {Collett}, T.~E. and {Acevedo Barroso}, J.~A. and others},
       title = "{Euclid Quick Data Release (Q1) The Strong Lensing Discovery Engine B -- Early strong lens candidates from visual inspection of high velocity dispersion galaxies}",
      journal = {A\&A, in press (Euclid Q1 SI), \url{https://doi.org/10.1051/0004-6361/202554605}},
         year = 2025,
        month = mar,
          eid = {arXiv:2503.15325},
        pages = {arXiv:2503.15325},
archivePrefix = {arXiv},
       eprint = {2503.15325},
 primaryClass = {astro-ph.GA},
       adsurl = {https://ui.adsabs.harvard.edu/abs/2025arXiv250315325E}
}

@ARTICLE{Q1-SP053,
       author = {{Euclid Collaboration: Lines}, N.~E.~P. and {Collett}, T.~E. and {Walmsley}, M. and others},
       title = "{Euclid Quick Data Release (Q1). The Strong Lensing Discovery Engine C -- Finding lenses with machine learning}",
      journal = {A\&A, in press (Euclid Q1 SI), \url{https://doi.org/10.1051/0004-6361/202554542}},
         year = 2025,
        month = mar,
          eid = {arXiv:2503.15326},
        pages = {arXiv:2503.15326},
archivePrefix = {arXiv},
       eprint = {2503.15326},
 primaryClass = {astro-ph.GA},
       adsurl = {https://ui.adsabs.harvard.edu/abs/2025arXiv250315326E}
}

@ARTICLE{Q1-SP054,
       author = {{Euclid Collaboration: Li}, T. and {Collett}, T.~E. and {Walmsley}, M. and others},
       title = "{Euclid Quick Data Release (Q1). The Strong Lensing Discovery Engine D -- Double-source-plane lens candidates}",
      journal = {A\&A, in press (Euclid Q1 SI), \url{https://doi.org/10.1051/0004-6361/202554543}},
         year = 2025,
        month = mar,
          eid = {arXiv:2503.15327},
        pages = {arXiv:2503.15327},
archivePrefix = {arXiv},
       eprint = {2503.15327},
 primaryClass = {astro-ph.GA},
       adsurl = {https://ui.adsabs.harvard.edu/abs/2025arXiv250315327E}
}

@ARTICLE{Q1-SP059,
       author = {{Euclid Collaboration: Holloway}, P. and {Verma}, A. and {Walmsley}, M. and others},
       title = "{Euclid Quick Data Release (Q1). The Strong Lensing Discovery Engine E - Ensemble classification of strong gravitational lenses: lessons for Data Release 1}",
      journal = {A\&A, submitted (Euclid Q1 SI)},
         year = 2025,
        month = mar,
          eid = {arXiv:2503.15328},
        pages = {arXiv:2503.15328},
archivePrefix = {arXiv},
       eprint = {2503.15328},
 primaryClass = {astro-ph.GA},
       adsurl = {https://ui.adsabs.harvard.edu/abs/2025arXiv250315328E}
}

@article{shajib_lens_review,
    author = {{Shajib, A.J., Vernardos, G., Collett, T.E.}},
    title = {Strong Lensing by Galaxies},
    journal = {Space Science Reviews},
    year = {2024},
    volume = {220},
    number={87}
}

@ARTICLE{Q1-TP001,
   author = {{Euclid Collaboration: Aussel}, H. and {Tereno}, I. and {Schirmer}, M. and others},
    title = "{Euclid Quick Data Release (Q1) - Data release overview}",
  journal = {A\&A, submitted (Euclid Q1 SI)},
     year = 2025,
    month = mar,
      eid = {arXiv:2503.15302},
    pages = {arXiv:2503.15302},
archivePrefix = {arXiv},
   eprint = {2503.15302},
primaryClass = {astro-ph.GA},
   adsurl = {https://ui.adsabs.harvard.edu/abs/2025arXiv250315302E}
}

@ARTICLE{Q1-SP047,
   author = {{Euclid Collaboration: Walmsley}, M. and {Huertas-Company}, M. and {Quilley}, L. and others},
    title = "{Euclid Quick Data Release (Q1): First visual morphology catalogue}",
  journal = {A\&A, accepted (Euclid Q1 SI)},
     year = 2025,
    month = mar,
      eid = {arXiv:2503.15310},
    pages = {arXiv:2503.15310},
archivePrefix = {arXiv},
   eprint = {2503.15310},
primaryClass = {astro-ph.GA},
   adsurl = {https://ui.adsabs.harvard.edu/abs/2025arXiv250315310E}
}

@misc{euclidcollaborationEuclidQuickData2025a,
   author = {{Euclid Collaboration: Quilley}, L. and {Damjanov}, I. and {de Lapparent}, V. and others},
    title = "{Euclid Quick Data Release (Q1). Exploring galaxy morphology across cosmic time through Sersic fits}",
  journal = {A\&A, in press (Euclid Q1 SI), \url{https://doi.org/10.1051/0004-6361/202554585}},
     year = 2025,
    month = mar,
      eid = {arXiv:2503.15309},
    pages = {arXiv:2503.15309},
archivePrefix = {arXiv},
   eprint = {2503.15309},
primaryClass = {astro-ph.GA},
   adsurl = {https://ui.adsabs.harvard.edu/abs/2025arXiv250315309E}
}

@article{Fischer2019,
	title = {{SDSS}-{IV} {MaNGA} {PyMorph} {Photometric} and {Deep} {Learning} {Morphological} catalogues and implications for bulge properties and stellar angular momentum},
	volume = {483},
	issn = {13652966},
	url = {https://academic.oup.com/mnras/advance-article/doi/10.1093/mnras/sty3135/5188692},
	doi = {10.1093/mnras/sty3135},
	number = {2},
	urldate = {2018-12-03},
	journal = {Monthly Notices of the Royal Astronomical Society},
	author = {Fischer, J. L. and Domínguez Sánchez, H. and Bernardi, M.},
	month = nov,
	year = {2019},
	pages = {2057--2077},
}

@article{Nair2010,
	title = {a {Catalog} of {Detailed} {Visual} {Morphological} {Classifications} for 14,034 {Galaxies} in the {Sloan} {Digital} {Sky} {Survey}},
	volume = {186},
	issn = {0067-0049},
	url = {http://arxiv.org/abs/1001.2401},
	doi = {10.1088/0067-0049/186/2/427},
	number = {2},
	journal = {The Astrophysical Journal Supplement Series},
	author = {Nair, Preethi B. and Abraham, Roberto G.},
	year = {2010},
	pages = {427--456},
}

@article{Baillard2011,
	title = {The {EFIGI} catalogue of 4458 nearby galaxies with detailed morphology},
	volume = {532},
	issn = {00046361},
	doi = {10.1051/0004-6361/201016423},
	number = {74},
	journal = {Astronomy and Astrophysics},
	author = {Baillard, A. and Bertin, E. and De Lapparent, V. and Fouqué, P. and Arnouts, S. and Mellier, Y. and Pelló, R. and Leborgne, J. F. and Prugniel, P. and Makarov, D. and Makarova, L. and McCracken, H. J. and Bijaoui, A. and Tasca, L.},
	year = {2011},
}
}

\newpage

\appendix

\section{Licensing}
\label{sec:licensing}

\subsection{Licensing for Existing Assets}
\label{sec:main_license}

Public astronomy data releases are intended for broad re-use by the research community. 
GZ2 uses Sloan Digital Sky Survey data released under a Creative Commons Attribution license\footnote{\href{https://www.sdss.org/collaboration/image-use-policy/}{https://www.sdss.org/collaboration/image-use-policy/}}
GZ Hubble and GZ CANDELS use images from the Hubble Space Telescope, which is operated by NASA. NASA images are not generally subject to copyright\footnote{\href{https://www.nasa.gov/nasa-brand-center/images-and-media/}{https://www.nasa.gov/nasa-brand-center/images-and-media/}}
GZ DESI (plus downstream datasets) uses Legacy Survey data published via NOIRLab. NOIRLab uses a Creative Commons Attribution 4.0 International License\footnote{\href{https://noirlab.edu/public/copyright/}{https://noirlab.edu/public/copyright/}}. 
GZ UKIDSS uses images from the UKIDSS survey; they do not include a specific license, but make clear that the data is intended is for open use\footnote{http://www.ukidss.org/archive/archive.html}.
GZ Euclid (downstream) uses images from the \textit{Euclid} space telescope, which is operated by ESA. ESA releases the images under a CC BY-NC 3.0 IGO license plus certain additional terms (e.g., no liability)\footnote{https://www.cosmos.esa.int/web/esdc/terms-and-conditions}. 

\subsection{Licensing for New Assets}

Galaxy Zoo volunteers freely contribute their time to advance science. It would inappropriate to exploit their efforts for commercial gain. We are therefore releasing the dataset with a non-commercial license (Creative Commons Attribution Non Commercial Share Alike 4.0).

We share the volunteer labels in the hope of advancing open foundation models. We therefore explicitly require that researchers training models on this dataset make the source code for constructing and training such models public by publication, such that other researchers can build upon their work.

\section{Core Dataset Subsets}
\label{sec:subsets}

This appendix provides a narrative summary of each subset of our Core dataset. Each subset asks volunteers a series of questions and answers (a decision tree). Each tree is visualized at \url{data.galaxyzoo.org/gz_trees/gz_trees.html}. These questions and answers aim to identify the features of a galaxy. Galaxies often have many features; for example, many galaxies with spiral arms also have bars. Features are better understood as attributes (e.g. this animal has four legs) rather than classes (e.g. this animal is a cat). The images used and the questions asked vary between subsets. This is ideal for building models that generalize to new images and new questions. We summarize these differences below, and refer the reader to the original astronomical publications for full details.

\begin{figure}
    \centering
    \includegraphics[width=\linewidth]{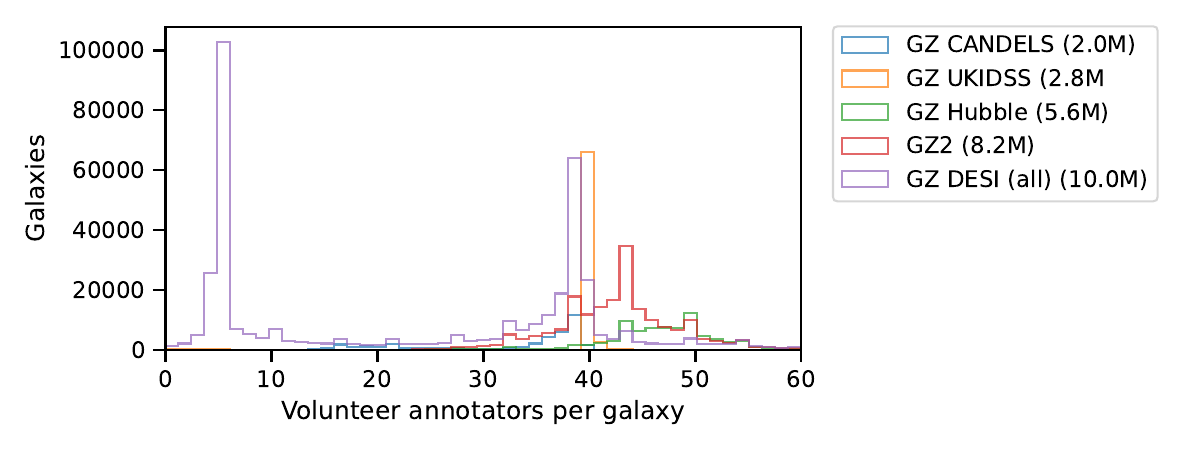}
    \caption{Distribution of the number of volunteer annotators per galaxy, per campaign. Most campaigns have approximately 40 annotators per galaxy. GZ DESI has a bimodal distribution with approximately half of galaxies receiving approximately 40 votes, and the remainder receiving approximately 5; see below. Legend shows the total number of completed decision tree annotations (i.e. the sum of the above) per campaign. }
    \label{fig:counts_per_campaign}
\end{figure}

Figure \ref{fig:counts_per_campaign} shows the distribution of the number of volunteer annotators per galaxy, per campaign. Across all campaigns (our Core dataset) volunteers completed 28.7M decision trees and 104M individual questions.

\subsection{Galaxy Zoo 2}

Galaxy Zoo 2 \cite{Willett2013} includes images from the Sloan Foundation Telescope in New Mexico \cite{Abazajian2009}. This was the first Galaxy Zoo campaign to use a decision tree; we will describe the other trees with reference to this tree. In short, volunteers are asked if a galaxy is smooth (i.e. a `blob'), featured (i.e. anything else of interest, typically a disk), or an artifact (i.e. an image problem). Volunteers selecting `smooth' are then asked to describe the shape of the galaxy and of any central `bulge' (i.e. a bright core). Volunteers selecting `featured' are instead asked to describe those features. These include spiral arms (swirling streams of stars, like milk in coffee), bars (a straight line of stars through the center of the galaxy), and bulges (above).

\subsection{Galaxy Zoo Hubble and Galaxy Zoo CANDELS}

Galaxy Zoo Hubble \cite{Willett2017a} and Galaxy Zoo CANDELS \cite{Simmons2017} use images from the Hubble Space Telescope. Hubble images are sharper than ground-based telescopes (i.e., the images have a narrower point-spread function than the effective ground-based point-spread function after accounting for atmospheric distortion). However, the galaxies in GZ Hubble and GZ CANDELS are more distant than in other campaigns, and so images in GZ Hubble and GZ CANDELS ultimately appear \textit{less} sharp.

\subsection{Galaxy Zoo DESI}

Galaxy Zoo DESI \cite{Walmsley2023desi} uses images primarily from the Blanco Telescope in Chile. GZ DESI is itself made up of four subcampaigns: GZD-1, GZD-2, GZD-5, and GZD-8. GZD-1 and GZD-2 used a question tree similar to GZ Hubble and GZ CANDELS, with only minor modifications to the number of possible answers to some questions (e.g. three vs. four possible bulge sizes). GZD-1 and GZD-2 ask identical questions on identically-processed images and so we jointly refer to them as GZD-1\&2. GZD-5 added a new question asking if galaxies are merging (this may be an important mechanism for galaxies to grow). GZD-5 also used an active learning strategy where volunteers were asked to provide 40 annotations for galaxies selected by the acquisition function (BALD, \cite{houlsby_bayesian_2011}) and 5 otherwise \cite{walmsley_galaxy_2020}. GZD-8 asked the same questions as GZD-5 for images taken from MzLS and BASS, two telescopes in the Northern Hemisphere. We present the class balance between various DESI campaigns grouped by decision tree (GZD-1/2, GZD-5, and GZD-8) in Fig. \ref{fig:class_label_distributions}.

\subsection{Galaxy Zoo UKIDSS}

GZ UKIDSS \cite{masters_galaxy_2024} includes images taken by the UK Infrared Telescope (UKIRT); the survey is described in \cite{Lawrence2007}.
Compared to images with optical wavelengths (e.g., those of GZ2 and GZ DESI), infrared images show more light from cooler sources -- primarily older stars. 
GZ UKIDSS used the same questions as GZ2.

\section{Comparison between Volunteer and Professional Astronomers}
\label{app:volunteer_and_professional}

\begin{figure}
    \centering
    \includegraphics[width=0.48\linewidth]{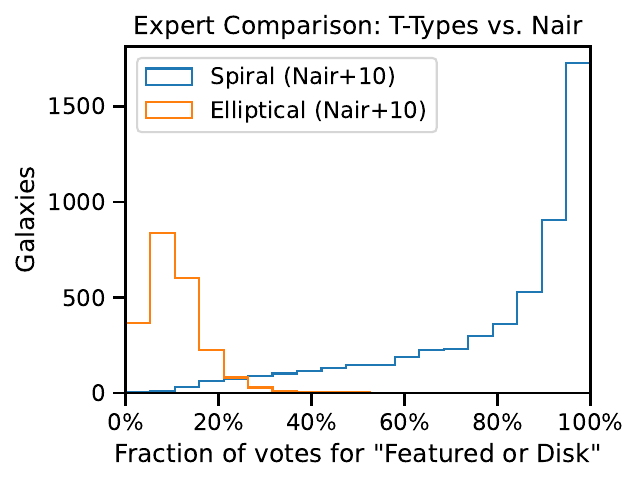}
    \hfill
    \includegraphics[width=0.48\linewidth]{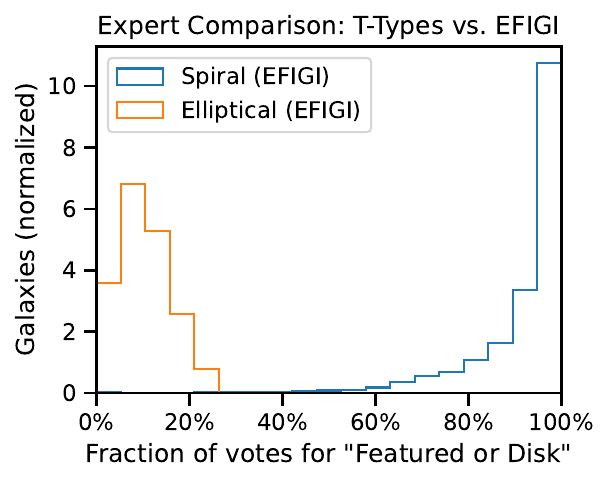}
    \caption{Agreement between GZ2 volunteers \cite{Willett2013} and professional astronomers (left: Nair et.~al.~\cite{Nair2010}, right: Baillard et.~al.~\cite{Baillard2011}). Agreement shown as the fraction of volunteers answering `Featured or Disk' to the question `Is this galaxy smooth, or featured?' vs. the professional label.}
    \label{fig:nair_efigi_feat}
\end{figure}

\begin{figure}
    \centering
    \includegraphics[width=0.48\linewidth]{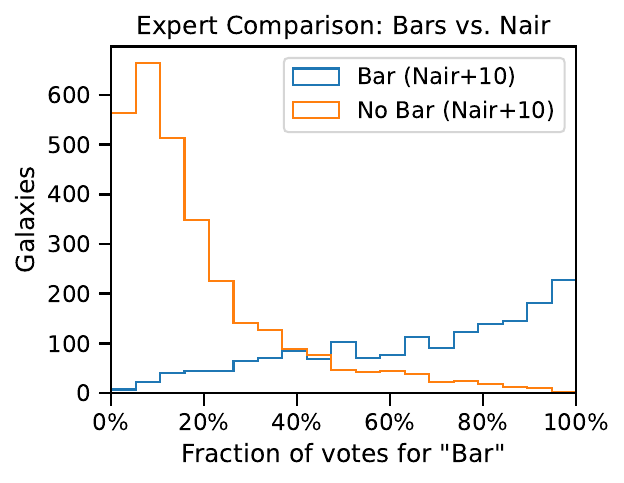}
    \hfill
    \includegraphics[width=0.48\linewidth]{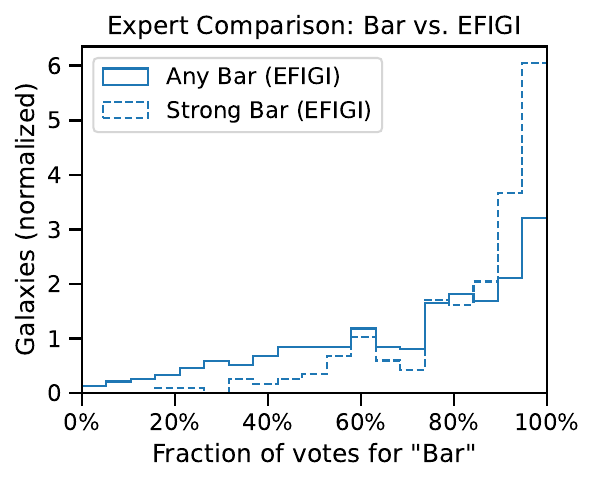}
    \caption{Agreement between GZ2 volunteers \cite{Willett2013} and professional astronomers (left: Nair et.~al.~\cite{Nair2010}, right: Baillard et.~al.~\cite{Baillard2011}). Agreement shown as the fraction of volunteers answering `Yes' to the question `Is there a bar' vs. the professional label. For Baillard et.~al., we group the professional annotations of (`Short', `Long', `Prominent') as `Any' and we group (`Long' `Prominent') as `Strong'. Volunteer vote fraction is a good proxy for both. Volunteers give higher vote fractions for `Strong' bars, as expected.}
    \label{fig:nair_efigi_bar}
\end{figure}

Surveys suggest that Galaxy Zoo volunteers are strongly motivated by a willingness to contribute to original scientific research \cite{raddick_galaxy_2013,jackson_unleashing_2024,jeong_how_2024}. 
The aggregated responses from GZ volunteers have been repeatedly shown to agree well with those of professional astronomers and with automated measurements \cite{Willett2013, Simmons2017, Fischer2019, dickinsonGalaxyZooClump2022, euclidcollaborationEuclidQuickData2025a}. Figures \ref{fig:nair_efigi_feat} and \ref{fig:nair_efigi_bar} show the agreement between Galaxy Zoo 2 aggregated volunteer responses and two independent teams of professional astronomers \cite{Nair2010,Baillard2011} for two questions answered by all three groups.

We can also compare the aggregated responses of volunteers in our GZ Rings downstream dataset with professional annotations from Buta et.~al.~\cite{Buta2017catalog}. 
3735 of 3840 (97.6\%) of galaxies annotated as ringed by Buta et.~al.~were identified as ringed by a majority of GZ volunteers.

\begin{figure}
    \centering
    \includegraphics[width=0.5\linewidth]{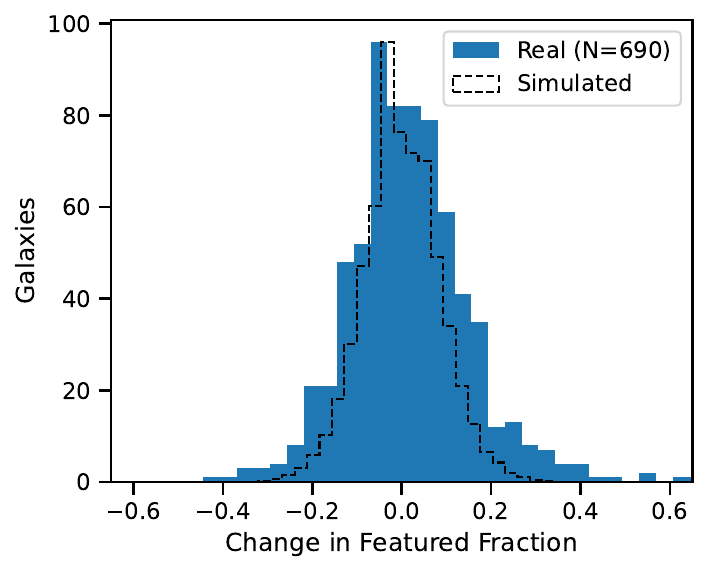}
    \caption{Change in `Featured' vote fraction for 690 galaxies accidentally uploaded twice, two years apart.  90\% of vote fractions changed by less than 20\%. This change is close to the irreducible aleatoric counting error expected when modeling each response as a biased coin flip.}
    \label{fig:gz_desi_repeat}
\end{figure}

Finally, we note that it is not necessary for volunteers to agree with professional annotators to agree for our dataset to be useful for training; we only need volunteers to be self-consistent. We estimate the consistency in our vote fractions by taking advantage of a natural experiment. During GZ DESI, 690 galaxies were accidentally uploaded twice, two years apart. Figure \ref{fig:gz_desi_repeat} shows the change in vote fraction for the `Featured' answer of the first decision tree question. 90\% of vote fractions changed by less than 20\%. We compare this with the change expected when simulating the volunteer responses as biased coin flips, using the observed vote fractions as the coin bias and tossing once per annotator. We find that the noise level in the aggregated responses is close to the irreducible aleatoric counting error.

\section{Datasheet}
\label{app:datasheet}

Each dataset is shared on the HuggingFace Hub with a concise practical summary. 
For completeness, we share a full datasheet below, following the `Datasheets for Datasets' framework \cite{gebru_datasheets_2021}. The datasheet answers show that Galaxy Zoo Evo is well-suited to open sharing for computer science research; it is an academic scientific dataset of galaxy images with no personal information, commercial motivations,  terrestrial implications, etc.

\subsection{Motivation}

\begin{enumerate}
    \item \textit{For what purpose was the dataset created?} The dataset was created to encourage development of new models for measuring the visual appearance of galaxies, as a benchmark for general-purpose foundation models, and as a real-world playground for open vision research problems e.g. efficient crowdsourcing, active learning, learning under label uncertainty, etc. 
    \item \textit{Who created the dataset and on behalf of which entity}? The dataset was compiled by Dr Mike Walmsley (MW) on behalf of the Galaxy Zoo collaboration. The collection of the original telescope images and volunteer labels was organised by the researchers named on the papers for each subset. The volunteer labels were contributed by members of the public, to whom we are deeply grateful.
    \item \textit{Who funded the creation of the dataset?} The dataset was funded by MW's postdoctoral fellowship at the University of Toronto (see Acknowledgements). The underlying data collection was funded by numerous grants over 15 years, listed individually in the papers for each subset. These grants were primarily from governments, academic institutions, and individual philanthropy (e.g. the Sloan Foundation).
\end{enumerate}

\subsection{Composition}

\begin{enumerate}
    \item \textit{What do the instances that comprise the dataset represent?} The instances are pairs of (galaxy, volunteer labels).
    \item \textit{How many instances are there in total?} There are 990k instances in total in this initial release (823k in Core and 167k in Downstream). We intend Galaxy Zoo Evo as a living dataset with new galaxy images and labels being collected and added.
    \item \textit{Does the dataset contain all possible instances or is it a sample of instances from a larger set?} Each GZ Evo subset contains all instances of galaxies imaged by a specific telescope operational campaign (a.k.a. `survey', in astronomical language) that meet campaign-specific criteria for relevance e.g. the galaxy must be bright and large enough to see detailed features, not suffer from obvious imaging artifacts, etc. 
    \item \textit{What data does each instance consist of?} The images are processed to represent the raw light (flux) collected by the telescope in an RGB format suitable for human viewing. The exact details vary by subset and are carefully documented in each associated paper. In short, the images have light collected by each telescope filter assigned to a channel (i.e. they are in false color) and use a compressed dynamic range to ensure faint features remain visible.
    \item \textit{Is there a label or target associated with each instance?} Each label is a vector of counts of volunteer answers to a series of questions (the Galaxy Zoo decision tree).
    \item \textit{Is any information missing from individual instances?} A small (less than 1\%) set of instances may suffer from image problems due to the challenging nature of taking telescope images of distant galaxies.
    \item \textit{Are relationships between individual instances made explicit?} The instances are related by position on the sky. This is shared as part of the dataset (right ascension and declination coordinates, analogous to latitude and longitude, measured in degrees).
    \item \textit{Are there recommended data splits?} Each subset has a canonical random train/test split, accessible via HuggingFace. The combined Galaxy Zoo Evo set has a train/test split consistent with the splits of each subset. 
    \item \textit{Are there any errors, sources of noise, or redundancies in the dataset?} As with all human annotations, the volunteer labels are imperfect and hence we expect label noise. See Sec. \ref{sec:limitations}.
    \item \textit{Is the dataset self-contained, or does it link to or otherwise rely on external resources?} The dataset is self-contained.
    \item \textit{Does the dataset contain data that might be considered confidential?} No.
    \item \textit{Does the dataset contain data that, if viewed directly, might be offensive, insulting, threatening, or might otherwise cause anxiety?} No. MW has found the galaxy images to be an effective antidote to anxiety.
\end{enumerate}

\subsection{Collection Process}

\begin{enumerate}
    \item \textit{How was the data associated with each instance acquired?} The images were acquired by professional astronomers using international telescope facilities. The labels were acquired primarily by presenting the images to volunteers at galaxyzoo.org.
    \item \textit{How were these mechanisms or procedures validated?} Publications for the underlying data have passed peer review by professional astronomers. See each associated paper for full validation details.
    \item \textit{Who was involved in the data collection process and how were they compensated?} Data annotation was primarily performed by volunteers at galaxyzoo.org. Volunteers were not compensated; they freely contributed their time to help advance our scientific understanding of the universe, for which we are grateful. Approximately 334k unique logged-in volunteers participated. 
    \item \textit{Over what timeframe was the data collected? Does this timeframe match the creation timeframe of the data associated with the instances?} The data was collected over a roughly 15-year period, and collection continues today. The timeframe associated with the instances is of the order 100 million to 10 billion years.
    \item \textit{Were any ethical review processes conducted (e.g., by an institutional review board)?} Our annotation approach was approved by an institutional review board at the University of Oxford, where the annotation platform was first developed.
\end{enumerate}

\subsection{Preprocessing/Cleaning/labelling}

\begin{enumerate}
    \item \textit{Was any preprocessing/cleaning/labeling of the data done?} The labels were acquired by presenting the images to online volunteers at galaxyzoo.org. 
    \item \textit{Was the “raw” data saved in addition to the preprocessed/cleaned/labeled data?} The original telescope images remain publicly available, although access by non-specialists may be challenging in some instances. See the associated paper for each subset for telescope details.
    \item \textit{Is the software that was used to preprocess/clean/label the data available?} The annotation platform is \href{zooniverse.org}{zooniverse.org} and the specific project is \href{galaxyzoo.org}{galaxyzoo.org}.
\end{enumerate}

\subsection{Uses}

\begin{enumerate}
    \item \textit{Has the dataset been used for any tasks already?} Galaxy Zoo labels have been used to support machine learning research since the 2014 Galaxy Challenge Kaggle competition, and remain the primary source of labelled data within this domain. An early internal version of the Galaxy Zoo Evo dataset has been used in a series of papers by the Galaxy Zoo collaboration investigating self-supervised learning and neural scaling laws for building foundation models. See Sec. \ref{sec:related_work}. Pre-release versions of this dataset were used for \cite{walmsley_scaling_2024} and \cite{pandyaSIDDASInkhornDynamic2025}.
    \item \textit{Is there a repository that links to any or all papers or systems that use the dataset?} No, as the dataset is not yet published.
    \item \textit{What (other) tasks could the dataset be used for?} We suggest research opportunities throughout the main text.
    \item \textit{Is there anything about the composition of the dataset or the way it was collected and preprocessed/cleaned/labeled that might impact future uses?} The GZ DESI subset used an active learning approach to select which galaxies to label and so is not a random set of possible galaxies.
    \item \textit{Are there tasks for which the dataset should not be used?} We recommend against using the dataset for precise benchmarking (in the style of e.g. `we achieve +0.1\% on ImageNet') as the volunteer labels have limited precision. In general, we feel very small accuracy improvements are less consequential for astronomy than methods addressing broader problems e.g. multi-task learning, transfer learning, uncertainty, etc.
\end{enumerate}

\subsection{Distribution}

Distribution (e.g. access, licensing, etc.) is addressed in the appendices above. We do not anticipate any regulatory restrictions.

\subsection{Maintenance}

\begin{enumerate}
    \item \textit{Who will be supporting/hosting/maintaining the dataset?} The dataset will be hosted by HuggingFace and maintained by the Galaxy Zoo Collaboration.
    \item \textit{How can the owner/curator/manager of the dataset be contacted (e.g., email address)?} Please email Dr Mike Walmsley using the contact details on \texttt{https://walmsley.dev/}.
    \item \textit{Will the dataset be updated?} Our hope is that the dataset will be updated as new images and labels become available. We anticipate updates on a frequency between several months and several years. Updates will be communicated via HuggingFace README pages. Updates are made on a best effort basis and conditional on continued funding for the collaboration.
    \item \textit{Will older versions of the dataset continue to be supported/hosted/maintained?} Older versions will be maintained where practical via HuggingFace versioning (extending git LFS).
    \item \textit{If others want to extend/augment/build on/contribute to the dataset, is there a mechanism for them to do so?} Contributions from astronomers with annotated galaxy images would be deeply welcome and credited appropriately. Please reach out to MW (above).
\end{enumerate}

\section{Reproducibility}

All benchmark results reported in this work can be reproduced using the GitHub repository \href{https://github.com/mwalmsley/gz-evo}{github.com/mwalmsley/gz-evo}. We include:

\begin{enumerate}
    \item A Dockerfile to ensure a consistent environment
    \item Utility code to download the Galaxy Zoo Evo dataset
    \item Scripts to run each baseline
    \item Configuration files documenting the hyperparameters used
\end{enumerate}

ConvNeXT-Nano can be reproduced on a consumer GPU. We recommend 2xA100-40GB GPUs or better for reproducing the larger models. 

\end{document}